\documentclass[11pt]{article}

\usepackage{amssymb}

\usepackage{graphicx}

\newfont{\ensmathquatorze}{msbm10 scaled 1400}
\newfont{\ensmathonze}{msbm10 scaled 1100}
\newfont{\ensmathdix}{msbm10}
\newfont{\ensmathneuf}{msbm10 scaled 833}
\newfont{\ensmathhuit}{msbm10 scaled 694}
\newfam\ensmathfam                        
\textfont\ensmathfam=\ensmathonze        
\scriptfont\ensmathfam=\ensmathdix       
\scriptscriptfont\ensmathfam=\ensmathhuit
\def\ensmf{\fam\ensmathfam\ensmathonze}         

\def\be{\begin{equation}}
\def\ee{\end{equation}}

\def\bea{\begin{eqnarray}}
\def\eea{\end{eqnarray}}
\def\beann{\begin{eqnarray*}}
\def\eeann{\end{eqnarray*}}

\renewcommand{\geq}{\geqslant}

\newcommand{\ket}[1]{|\kern.3ex#1\kern.3ex\rangle}
\newcommand{\bra}[1]{\langle\kern.3ex #1 \kern.3ex|}
\newcommand{\APPROX}[1]{                
   {{\raisebox{-.3cm}{$\textstyle\simeq$}} \atop {\scriptstyle{#1}}}}

\newcommand{\mean}[1]{\left\langle #1 \right\rangle} 
\newcommand{\smean}[1]{\langle #1 \rangle} 

\newcommand{\EXP}[1]{{\mbox{\large e}}^{#1}}         
\renewcommand{\cosh}[1]{\mathop{\mathrm{ch}}\nolimits #1} 

\newcommand{\im}{\mathop{\mathrm{Im}}\nolimits}      
\newcommand{\tr}[1]{\mathop{\mathrm{Tr}}\nolimits\left\{ #1 \right\}}  
\newcommand{\sign}{\mathop{\mathrm{sign}}\nolimits}  

\def\NN{{\ensmf N}}                 
\def\RR{{\ensmf R}}                 

\def\I{{\rm i}}                  
\def\D{{\rm d}}                  
\def\Dc{{\rm D}}                 

\newcommand{\drond}[2]{\frac{\partial #1}{\partial #2}} 

\newcommand\ab{{\alpha\beta}}
\newcommand\ba{{\beta\alpha}}

\begin{document}

\title{Charge and currents distribution in graphs}

\author{Christophe Texier$^{(a)}$ and Pascal Degiovanni$^{(b)}$}
 
\date{\today} 

\maketitle	

\hspace{2cm}
\begin{minipage}[t]{12cm}
{\small
$^{(a)}$Laboratoire de Physique Th\'eorique et Mod\`eles Statistiques.

Universit\'e Paris-Sud,
B\^at. 100, F-91405 Orsay Cedex, France.

\hspace{3cm}and

Laboratoire de Physique des Solides.

Universit\'e Paris-Sud,
B\^at. 510, F-91405 Orsay Cedex, France.

\vspace{0.5cm}

$^{(b)}$Laboratoire de Physique de l'ENS de Lyon (UMR CNRS 5672).

\'Ecole Normale Sup\'erieure de Lyon.

46, all\'ee d'Italie, 69364 Lyon Cedex 07, France.
}
\end{minipage}

\begin{abstract}
We consider graphs made of one-dimensional wires connected at vertices, and 
on which may live a scalar potential. We are interested in a scattering 
situation where such a network is connected to infinite leads.
We study the correlations of the charge in such graphs out 
of equilibrium, as well as the distribution of the currents in the wires, 
inside the graph.
These quantities are related to the scattering matrix of the graph.
We discuss the case where the graph is weakly connected to the wires.
\end{abstract}

\vspace{0.5cm}

\noindent
PACS~: 03.65.Nk, 73.23.-b





\section{Introduction\label{sec:Intro}}

Within the field of mesoscopic physics, the interest in graphs  
is motivated by the fact that
they provide simple models for networks of wires, which are most of the time 
sufficient to describe the effect of interest (such as Aharonov-Bohm 
oscillations of the conductance of a ring for example). 
Scattering theory plays a central role in mesoscopic physics~: it provides
a transparent formalism to study transport properties of phase coherent
systems. 
Moreover, many other physical quantities can be related to scattering 
properties, like the current noise \cite{Les89,But90a}, the density of states
through the Friedel sum rule, mesoscopic capacitance, relaxation resistance 
\cite{But93,ButPreTho93}. 
Scattering on graphs has attracted the attention of many authors among
which we can quote
\cite{GerPav88,AvrSad91,Ada92,KosSch99,KotSmi00,TexMon01,Tex02,KotSmi02,TexBut03,BarGas01}.

Despite the scattering matrix is a global quantity characterizing the
full system, some local information can be extracted from it. This idea
has been fruitfully exploited in many works of B\"uttiker {\it et al}
(see review articles \cite{But00,But02} and references therein).
To understand this point
let us first consider the case of an isolated system (an isolated graph for
example). In this case, the spectrum of the Schr\"odinger operator is
discrete~: $E_n$, $\varphi_n(x)$. Let us consider a physical quantity
described by the operator $\hat X$, related to a conjugate variable $f$
(that is $\hat X=\frac{\partial}{\partial f}\hat H$ where $\hat H$ is
the Hamiltonian).
Typical examples are provided by a magnetization ${\cal M}$, a persistent
current $I$, or the local density $\rho(x)$~, which are conjugated to
the magnetic field $-{\cal B}$, a flux\footnote{
  The variable conjugate to a flux line threading a loop
  of a planar graph is the current flowing through the semi infinite line 
  issuing
  from the flux \cite{AkkAueAvrSha91} (see also \cite{ComMorOuv95}). 
  This can also be easily understood in the 2-dimensional plane 
  \cite{DesOuvTex98}.
} $-\phi$ and the potential $V(x)$, respectively. 
As it is well known, a simple way to obtain the expectation
value of the physical quantity of interest is to compute the derivative of 
the eigenenergies with respect to $f$~:
$X_n=\bra{\varphi_n}\hat X\ket{\varphi_n}=\frac{\partial E_n}{\partial f}$.
This result is known as the Feynman-Hellmann theorem.

A natural question is to extend these relations to open
systems that are connected to reservoirs possibly in an out of
equilibrium situation. As it is well known, the scattering approach
will prove to be relevant for this purpose. The open system of
interest in the present article will be a graph connected to some
infinite wires. In this case the spectrum is continuous. The
stationary scattering states $\tilde\psi^{(\alpha)}_E(x)$
describing the injection of a plane wave at contact $\alpha$
provide the convenient basis of states for the discussion. Then we
can relate the quantity of interest (that can give a local
information) to the scattering matrix $\Sigma$ through the
relation: $\bra{\tilde\psi^{(\alpha)}_E}\hat
X\ket{\tilde\psi^{(\beta)}_E}= -\frac{1}{2\I\pi}
\left(\Sigma^\dagger\frac{\partial\Sigma}{\partial
f}\right)_{\alpha\beta}$.
In the case of graphs, this idea will be made explicit in three
cases: ({\it i}) when $\hat X\to\hat\rho(x)$ is the local density
of electrons\footnote{In this case
  $\bra{\tilde\psi^{(\alpha)}_E}\hat\rho(x)\ket{\tilde\psi^{(\beta)}_E}
  =\tilde\psi^{(\alpha)*}_E(x)\tilde\psi^{(\beta)}_E(x)$ is an off
  diagonal element of the local DoS \cite{But00,TexBut03} 
}
(then $f\to V(x)$ is the local potential), ({\it ii}) if $\hat X\to\hat
Q$ is the charge of the graph ($f\to U$ is a potential constant
inside the graph) and ({\it iii}) if $\hat X\to\hat J_{\mu\nu}$ is
the current in an arc of the graph ($f\to \theta_{\mu\nu}$ is the
flux along the arc).

\medskip

The purpose of our article is to study the 
distribution of charge and current densities in a graph out of
equilibrium. The out of equilibrium regime is obtained by imposing
different potentials at the external leads. A motivation for this
study comes from the recent interest in quantum coherent devices
such as Cooper pair boxes used for building charge Qubits (see
\cite{MakSchShn01} for a review). The full spectrum of 
charge fluctuations is involved in the study of the dephasing in a
Qubit perturbed by the charge fluctuations of another conductor
capacitively coupled to the first one
\cite{But00,PilBut02,CleGirSto03}. In the same way, current
density fluctuations are a source of dephasing for Qubits based on
flux states. Since the formalism developed in the present paper
provides a systematic way for evaluating the charge and current
density noise fluctuations in a mesoscopic circuit, it might be
useful for estimating the dissipation and decoherence properties of some
experimental systems of Qubits. More precisely, it was shown  that transition 
rates of a two level system weakly coupled to a quantum environment are 
directly 
related to the unsymmetrized correlator~: see for example 
\cite{SchCleGirLehDev02} where the roles of the negative and the positive
part of the spectrum of the unsymmetrized correlator are studied.
On the other hand the relaxation and decoherence rates are related to a 
symmetrized correlator \cite{MakSchShn01}.
Correlators are more directly accessible in noise measurements~:
in a recent work, Gavish {\it et al.} \cite{GavLevImr00,GavImrLevYur02} 
proposed a description of the full measurement chain for the current noise 
of a mesoscopic sample. In this work, the unsymmetrized correlator is 
involved in excess noise measurement.
Finally one should mention that experimentalists are now able, using photon 
assisted tunneling in a superconductor-insulator-superconductor tunnel 
junction, to measure the unsymmetrized current noise correlator in quantum 
mesoscopic devices \cite{DebOnaGurKou03}.
The question of which correlator (symmetrized or not) to consider 
depends on the question of interest.
Therefore we will consider in the following the unsymmetrized correlator as 
the fundamental object\footnote{
  There are indeed two unsymmetrized correlators depending on 
  the order chosen. But for a system in a stationary regime, they can be 
  simply related as shown in appendix \ref{app:unsymmetrized}.
}.

In this paper, electron-electron interactions will not be taken into 
account. However, even if this limits the applicability of our results, 
we recall that they can be taken into account in a mean field Hartree
approximation within the scattering approach, as it has been developed in
several papers by B\"uttiker and collaborators 
\cite{But93,ButPreTho93,ButThoPre94} (see also \cite{ButChr97} for a review). 
In this framework the charge (or the current) contains two contributions~: 
a bare contribution (injected charge) and a contribution from screening. 
Screening affects ac transport or finite frequency noise.
It is not the purpose of our article to consider such interaction effects; 
in other terms we will focus on the bare contribution of the charge 
and its relation to the scattering matrix of a graph.

This paper is organized as follows: first of all, the basic
formalism \cite{TexMon01} necessary for the discussion is
recalled. Then the charge distribution inside the graph is
analyzed in details. The first and second moments of the total
charge are related to the scattering matrix. Finally analytic
expressions for the full spectrum of charge fluctuations are
provided.

\medskip

In a second part we will show how currents inside the graph can be
related to the scattering matrix. It was shown in \cite{AkkAueAvrSha91} 
that the persistent current can be related to the derivative of the Friedel 
phase with respect to the magnetic flux. In this work the possible 
generalization to an out of equilibrium situation was not considered because
the authors did not identify the different contributions of the various 
scattering states associated to the different leads.
These contributions were identified later in \cite{Tan01}. 
Moreover, in Taniguchi's work, a formula relating the current-current 
correlations and the scattering matrix was proposed. 
Despite the contributions of the scattering
states to the correlator were given, it is still not sufficient to 
study current-current correlations in an out-of-equilibrium situation
(this point will be made clear later). 
Our results go beyond this limitation and provide the generalization of 
Taniguchi's result.


\section{Basic formalism~: scattering matrix\label{sec:Forma}}

We consider the Schr\"odinger operator $-\Dc_x^2+V(x)$ on a graph, where 
$\Dc_x=\D_x-\I A(x)$ is the covariant derivative 
(we choose units ~: $\hbar=2m=e=1$).
The graph is made of $B$ bonds $(\ab)$, each being identified with an interval
$[0,l_\ab]\in\RR$. We call $x_\ab$ the coordinate that measures the distance
from the vertex $\alpha$.
The Schr\"odinger operator acts on a scalar function $\varphi(x)$ which is 
described by $B$ components $\varphi_{(\ab)}(x_\ab)$, one for each bond.
The bonds are connected at $V$ vertices. The adjacency matrix $a_\ab$
encodes the structure of the graph~: $a_\ab=1$ if $(\ab)$ is a bond and 
$a_\ab=0$ otherwise.

\subsection*{Vertex formulation}

Let us first assume that the wave function 
is continuous at each vertex. This allows to introduce vertex variables~;
we denote $\varphi_\alpha\equiv\varphi(\alpha)$ the function at the vertex 
$\alpha$. The continuity condition reads~: 
$\varphi_{(\ab)}(x_\ab=0)=\varphi_\alpha$ for all vertices $\beta$ 
neighbours of $\alpha$.
A second condition is added to ensure current conservation at the vertices~:
$\sum_\beta a_\ab\Dc_x\varphi_{(\ab)}(x_\ab=0)=\lambda_\alpha\varphi_\alpha$ 
where
the sum over $\beta$ runs over all neighbouring vertices of $\alpha$ due to
the presence of the adjacency matrix. $\lambda_\alpha$ is a real parameter.
The requirement of continuity of the wave function imposes a special
scattering at the vertices~: in particular, the transmission amplitudes
of a plane wave of energy $E=k^2$ between two leads issuing from the same 
vertex of coordinence $m_\alpha=\sum_\beta a_\ab$ are all equal to 
$2/(m_\alpha+\I\lambda_\alpha/k)$.

The wave function on the bond $(\ab)$ is~:
\be
\varphi_{(\ab)}(x_\ab) = \EXP{\I x_\ab\theta_\ab/l_\ab }
\left( \varphi_\alpha\, f_\ab(x_\ab)  
     + \varphi_\beta\, \EXP{-\I\theta_\ab} f_\ba(x_\ab) \right)
\ee
where $\theta_\ab$ is the magnetic flux along the bond $(\alpha\beta)$
(the vector potential is $A_\ab=\theta_\ab/l_\ab$). The two real functions
$f_\ab(x)$, $f_\ba(x)$ are the two linearly independent solutions of the 
Schr\"odinger equation $[E+\D_x^2-V_{(\ab)}(x)]f(x)=0$
on the bond satisfying boundary conditions~:
$f_\ab(0)=1$, $f_\ab(l_\ab)=0$, $f_\ba(0)=0$ and $f_\ba(l_\ab)=1$. 
These two functions encode the information about the potential on the bond.
For example, in the absence of potential, $V(x)=0$, we have
$f_\ab(x_\ab)=\frac{\sin k(l_\ab-x_\ab)}{\sin kl_\ab}$.

\begin{figure}[!ht]
\begin{center}
\includegraphics[scale=1]{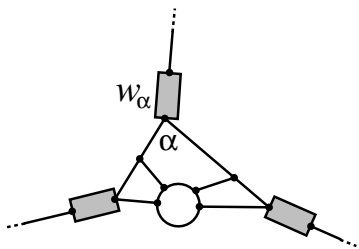}
\end{center}
\caption{Example of graph. The boxes represent the couplings between the 
         infinite leads and the graph.\label{fig:agraph}}
\end{figure}

The graph is connected to $L$ leads. Each lead is a semi-infinite line
plugged at a vertex of the graph, with a coupling parameter
$w_\alpha\in\RR$ (see figure \ref{fig:agraph}). 
The introduction of these couplings allows to go continuously
from an isolated graph to a connected one. The precise physical meaning
of these parameters is given in \cite{TexMon01}. In particular, the 
transmission amplitude through the box between the graph and the lead is 
$2w_\alpha/(1+w_\alpha^2)$.
We introduce the $L\times V$ matrix $W$~:
\be
W_\ab = w_\alpha\,\delta_\ab
\ee
where $\alpha$ belongs to the set of vertices connected to leads and 
$\beta$ to the set of all vertices of the graph. This matrix encodes
the information about the way the graph is connected to the external leads.

The scattering matrix $\Sigma$ is a $L\times L$ matrix describing how a 
plane wave of energy $E$ entering from a lead is scattered into the other 
leads by the graph. It is given by 
\be\label{RES2}
\Sigma = -1 + 2\,W\frac{1}{M+W^{\rm T}W}W^{\rm T}
\ee
where the matrix $M$ is~:
\be\label{MJean}
M_\ab(-E) = \frac{\I}{\sqrt{E}}
\left(
  \delta_\ab
  \left[
    \lambda_\alpha - 
    \sum_\mu a_{\alpha\mu} \frac{\D f_{\alpha\mu}}{\D x_{\alpha\mu}}(\alpha)
  \right]
  + a_\ab \frac{\D f_\ab}{\D x_\ab}(\beta)\,\EXP{\I\theta_\ab}
\right)
\:.\ee
Note that for $E>0$, this matrix is antihermitian~: $M^\dagger=-M$.
It can also be related \cite{TexMon01} to reflexion and transmission 
coefficients describing the potential on each bond~:
\bea\label{RES3}
M_\ab(-E) &=&
\delta_\ab\left(\I\frac{\lambda_\alpha}{\sqrt{E}}
+ \sum_\mu a_{\alpha\mu}
\frac{(1-r_{\alpha\mu})(1+r_{\mu\alpha})+t_{\alpha\mu}\,t_{\mu\alpha}}
     {(1+r_{\alpha\mu})(1+r_{\mu\alpha})-t_{\alpha\mu}\,t_{\mu\alpha}}
\right)
\nonumber \\ && \hspace{1cm}
- a_\ab\frac{2\,t_\ab}{(1+r_\ab)(1+r_\ba)-t_\ab\,t_\ba}
\:.\eea
The expressions (\ref{MJean},\ref{RES3}), together with (\ref{RES2}), 
generalize results known in the absence of the potential \cite{AvrSad91}.

The vertex formulation that we have just recalled is rather efficient
mainly because vertex matrices are rather compact. 
However, as we have noticed, it corresponds to a particular choice of 
vertex scattering which does not describe all allowed relevant physical
situations.
In the most general situation it is not anymore possible to introduce vertex
variables and one has to use the arc formulation that will now be briefly 
described.

\subsection*{Arc formulation}

\begin{figure}[!ht]
\begin{center}
\includegraphics{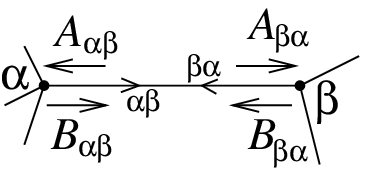}
\end{center}
\caption{The internal amplitudes associated to the arcs $\ab$ and $\ba$.
         \label{fig:arcs}}
\end{figure}

An arc is an oriented bond. On each arc $i$ we introduce an amplitude 
$A_i$ arriving at the vertex from which $i$ issues and an amplitude 
$B_i$ departing from it (see figure \ref{fig:arcs}).
Equivalently, the wave function $\psi_i(x)$ on the bond is matched with 
$A_i\EXP{-\I kx}+B_i\EXP{\I kx}$ at the extremity of the arc.
It is clear than we have to introduce $L$ such couples of amplitudes, one
for each external lead. These external amplitudes are gathered in $L$-column
vectors $A^{\rm ext}$ and $B^{\rm ext}$. By definition the scattering matrix 
relates these amplitudes~: $B^{\rm ext}=\Sigma A^{\rm ext}$.
On the other hand we must introduce two couples of amplitudes $A_i$, $B_i$
per bond of the graph, {\it i.e.} one couple per arc. We gather these $2B$
amplitudes into the column vectors $A^{\rm int}$ and $B^{\rm int}$.
Finally we group all amplitudes, internal and external, in two 
$2B+L$ column vectors $A$ and $B$.

The scattering by the bonds is described by a matrix
$R$ coupling reversed internal arcs~: $A^{\rm int}=RB^{\rm int}$. The 
matrix element between arc $i$ and $j$ is given by~:
\be
R_{ij} = r_i \delta_{i,j} + t_{\bar i} \delta_{\bar i,j}
\ee
where $\bar i$ designate the reversed arc. $r_i$ and $t_i$ are the 
reflexion and transmission coefficients describing the scattering
of a plane wave by the potential of the bond $(i)$.
The scattering at the vertices is described by a matrix $Q$ coupling 
arcs issuing from the same vertex~: $B=QA$.
If the basis of arcs is organized as 
$\{\mbox{internal arcs},\mbox{external arcs}\}$, the matrix $Q$ can be 
separated into blocks:
\be\label{vs}
Q = 
\left(
\begin{array}{c|c}
Q^{\rm int} & \tilde Q^{\rm T} \\ \hline 
\tilde Q   & Q^{\rm ext}
\end{array}
\right)
\ee
The scattering matrix reads~:
\be\label{RES1}
\Sigma = Q^{\rm ext} + 
\tilde Q \, ({R^{\dagger} - Q^{\rm int}})^{-1} \, \tilde Q^{\rm T}
\:.\ee
For more details, see \cite{TexMon01}. 
Historically, the arc approach has been followed in many works, like
\cite{ButImrLan83,GefImrAzb84,ButImrAzb84} since it is the most natural
approach. It has been formalized more systematically in \cite{KotSmi99}
without potential and in \cite{TexMon01} in the most general case.

How can we express the wave function inside the graph within the arc
formulation~?
In this case, the appropriate basis of solutions of the Schr\"odinger 
equation on the bond $[E+\D_x^2 - V_{(\ab)}(x)]f(x)=0$ is not anymore the 
functions  $f_\ab(x)$ and $f_\ba(x)$ introduced above, but the couple of 
stationary scattering states $\phi_\ab(x)$ and $\phi_\ba(x)$ associated 
to the potential $V_{(\ab)}(x)$ on the bond $(\ab)$.
If we imagine that the potential $V_{(\ab)}(x)$ is embedded in $\RR$, then
the function $\phi_\ab(x)$ is the scattering state incoming on the potential
from the vertex $\alpha$ and is matched out of the bond to~:
$\phi_\ab(x)=\EXP{\I kx}+r_\ab\EXP{-\I kx}$ for $x<0$ and 
$\phi_\ab(x)=t_\ab\EXP{\I k(x-l_\ab)}$ for $x>l_\ab$ \cite{TexMon01}.

Then the component of the wave function $\varphi(x)$ on the bond $(\ab)$ 
reads~:
\be
\varphi_{(\ab)}(x_\ab) = B_\ab\,\phi_\ab(x_\ab)+B_\ba\,\phi_\ba(x_\ab)
\:.\ee


\section{Charge of the graph}

This section is devoted to the study of the charge distribution in
the graph. Our discussion will focus on the average and
correlation of the total charge of the graph. The average
charge and the zero frequency charge noise can be related to
the graph's scattering matrix. These relations provide an extension
of the Feynman-Hellman theorem for open systems in an out of equilibrium 
situation.
Then, we present a detailed study of the
charge noise at finite frequency, emphasizing the effect of the
non equilibrium regime. For simplicity, we shall work at vanishing
temperature.

\medskip

It is convenient to use the language of ``second-quantization'' and introduce
the field operator~:
\be
\hat\psi(x,t) = \sum_{\alpha=1}^L\int_0^\infty\D E\:
\tilde\psi^{(\alpha)}_E(x)\,\hat a_\alpha(E)\,\EXP{-\I Et}
\:,\ee
where $\hat a_\alpha(E)$ is the annihilation operator associated to the 
stationary scattering state $\tilde\psi^{(\alpha)}_E(x)$ corresponding to
a plane wave of energy $E$ injected from the lead $\alpha$. 
Note that 
$\hat\psi(x,0)\ket{\tilde\psi^{(\alpha)}_E}
=\tilde\psi^{(\alpha)}_E(x)\ket{\rm vacuum}$.

Studying the charge distribution for graphs with localized states
\cite{Rot84,Tex02,TexBut03} would require taking into account the
contribution of the discrete spectrum in the field operator~:
$\sum_n\sum_{j=1}^{g_n}\varphi_{n,j}(x)\,\hat a_{n,j}\,\EXP{-\I E_nt}$
(the function $\varphi_{n,j}(x)$ is an eigenstate of energy $E_n$ 
localized in the graph and thus normalized to unity in the graph, and $j$ 
denotes a degeneracy label). However, such a situation
in not generic but arises from symmetries of the graph. For this
reason, we shall not discuss it here.

In a non equilibrium situation, the quantum statistical average gives~:
$\smean{{\hat a}_\alpha^\dagger(E)\hat a_\beta(E')}
=\delta_\ab\delta(E-E')\,f_\alpha(E)$ where $f_\alpha(E)$ is the Fermi-Dirac
distribution function giving the occupation of the scattering states coming from 
the lead $\alpha$.

\medskip

\noindent{\bf Charge operator.}
The charge operator is~:
\be
\hat Q(t) = \int_{\rm Graph}\D x\,\hat\psi^\dagger(x,t)\hat\psi(x,t)
\:.\ee
We introduce its matrix elements on the shell of energy $E$~:
\be
\rho^{(\alpha,\beta)}(E)=
\bra{\tilde\psi^{(\alpha)}_E}\hat Q(t)\ket{\tilde\psi^{(\beta)}_E}
\:.\ee
Since the spectrum is continuous, these matrix elements have the dimension of 
a density of states (DoS). They can be related to the scattering matrix~:
\be
\rho^{(\alpha,\beta)}(E)=
-\frac1{2\I\pi}\left( \Sigma^\dagger\frac{\D\Sigma}{\D U}\right)_\ab
\:,\ee
where $U$ is a constant potential added inside the graph only (the variable 
conjugate to the charge of the graph).
Instead of differentiating with respect to some additional background 
potential $U$, it is also possible to relate it to the derivative with respect
to the energy~:
\be\label{Smith}
\rho^{(\alpha,\beta)}(E)=\int_{\rm Graph}\D x\,
\tilde\psi^{(\alpha)}_{E}(x)^*\,\tilde\psi^{(\beta)}_{E}(x)
=
\frac1{2\I\pi}
\left(
  \Sigma^\dagger\frac{\D\Sigma}{\D E}
  +\frac{1}{4E}(\Sigma-\Sigma^\dagger)
\right)_\ab
\:.\ee
Note that $\sum_\alpha\rho^{(\alpha,\alpha)}(E)$ is the 
DoS of the graph, {\it i.e.} the local DoS integrated inside the graph.
These relations are proven in appendix \ref{app:Smith}. 

\medskip

\noindent{\bf Average charge.}
The average charge~:
\be\label{qmoy}
\smean{\hat Q(t)} = \sum_\alpha\int_0^\infty\D E\,f_\alpha(E)\,
\int_{\rm Graph}\D x\,|\tilde\psi^{(\alpha)}_E(x)|^2
= \sum_\alpha\int_0^\infty\D E\,f_\alpha(E)\,\rho^{(\alpha,\alpha)}(E)
\ee
involves the injectivities 
$\rho(x,\alpha;E)=|\tilde\psi^{(\alpha)}_E(x)|^2$, which are the contributions
to the local density of states (LDoS) coming from the scattering state 
$\tilde\psi^{(\alpha)}_E(x)$.
The contribution $\rho^{(\alpha,\alpha)}(E)$ of the scattering state
$\tilde\psi^{(\alpha)}_E$ to the DoS of the graph is weighted by the 
occupation Fermi factor in the lead $\alpha$.
This illustrates the necessity of the concept of injectivities, emissivities, 
etc, in the context of out of equilibrium systems \cite{But93,GasChrBut96} 
(see also \cite{TexBut03} for a discussion in the context of graphs).

\medskip

\noindent{\bf Charge correlation function.}
The charge correlation function is defined as~:
\be\label{corrdef}
S_{QQ}(\omega) = \int_{-\infty}^{+\infty}\D(t-t')\,
\left(
  \smean{\hat Q(t)\,\hat Q(t')} - \smean{\hat Q(t)}\smean{\hat Q(t')}
\right)\,\EXP{\I\omega(t-t')}
\:.\ee
In appendix \ref{app:unsymmetrized}, the relation between this 
unsymmetrized correlator and the other one ($Q(t)$ and $Q(t')$ in 
reverse order) is clarified. Definition (\ref{corrdef}) matches with the 
one used reference \cite{SchCleGirLehDev02} which contains a detailed 
discussion of the relation between the unsymmetrized correlator and
transition rates in a two level system linearily coupled to the charge 
operator.

Using the relation
$
\smean{
  {\hat a}^\dagger_\alpha{\hat a}_\beta{\hat a}^\dagger_\mu{\hat a}_\nu}
- \smean{{\hat a}^\dagger_\alpha{\hat a}_\beta}
  \smean{{\hat a}^\dagger_\mu{\hat a}_\nu}
=\delta_{\alpha\nu}\,\delta_{\beta\mu}\, f_\alpha(1-f_\beta)
$
we obtain~:
\bea\label{qfluc}
S_{QQ}(\omega) = 2\pi\sum_{\alpha,\beta}\int_0^\infty\D E\,
f_\alpha(E)[1-f_\beta(E+\omega)] 
\left|
  \int_{\rm Graph}\D x\:
  \tilde\psi^{(\alpha)}_{E}(x)^*\, \tilde\psi^{(\beta)}_{E+\omega}(x)\,
\right|^2
\:.\eea
Only the zero frequency correlations involve the $\rho^{(\alpha,\beta)}(E)$~:
\be
S_{QQ}(\omega=0) = 2\pi\sum_{\alpha,\beta}\int_0^\infty\D E\,
f_\alpha(E)[1-f_\beta(E)]\,
\rho^{(\alpha,\beta)}(E)\,\rho^{(\beta,\alpha)}(E)
\:.\ee
Using equation (\ref{Smith}) the zero frequency noise of the total charge
is related to the scattering matrix of the graph.

\medskip

\noindent{\bf Charge fluctuations.} 
The charge fluctuations at a given time involve the integral of the full 
spectrum~:
\be\label{cf1}
q_2 = \smean{\hat Q(t)^2} - \smean{\hat Q(t)}^2
= \frac{1}{2\pi}\int\D\omega\,S_{QQ}(\omega)
\:.\ee
In terms of the stationary scattering states, we get~:
\be\label{cf2}
q_2 = \sum_{\alpha,\beta}\int_0^\infty\D E\D E'\,
f_\alpha(E)[1-f_\beta(E')]\,
\left|
  \int_{\rm Graph}\D x\:
  \tilde\psi^{(\alpha)}_{E}(x)^*\, \tilde\psi^{(\beta)}_{E'}(x)\,
\right|^2
\:.\ee


\subsection*{Weakly connected graphs}

To go further let us focus on the case of graphs weakly coupled to the 
leads ($w_\alpha\to0$). Note that we do not consider charging effect in the 
following (Coulomb blockade) which is important if the capacitance 
describing the Coulomb interaction between the leads and the graph 
is small (see \cite{KouMarEueTarWesWin97} for a review article). A 
description of such effects would require a different approach.
However, in the neighbourhood of the Coulomb peak, a description within the
scattering approach can be sufficient to describe transport, like it has been
done very recently in \cite{KobAikKatIye03} to analyze Fano profile 
measurements in a ring with a dot embedded in one of its arm.

If $w_\alpha\to0$ the decomposition of the scattering states over the 
resonances (levels of the isolated graph), derived
in appendix \ref{app:resonance}, can be used
\be\label{rwf}
\tilde\psi^{(\alpha)}_{E}(x) \simeq
\sum_n 
\frac{1}{\sqrt{\pi}}\,
\frac{\I E_n^{1/4} w_\alpha \varphi_n^*(\alpha)}{E-E_n+\I\Gamma_n}\,
\varphi_n(x)
\:.\ee
Here $\varphi_n(x)$ denotes the wave function of the eigenstate of energy 
$E_n$ of the isolated graph, normalized to unity in the graph.
From this expression we get~:
\be
\rho^{(\alpha,\beta)}(E) \simeq \sum_n \frac{1}{\pi} 
\frac{\sqrt{\Gamma_{n,\alpha}\Gamma_{n,\beta}}\,\EXP{\I\chi_\ab}}
     {(E-E_n)^2+\Gamma_n^2}
\:,\ee
where $\Gamma_{n,\alpha}=\sqrt{E_n}w_\alpha^2|\varphi_n(\alpha)|^2$ is the
contribution of the contact $\alpha$ to the resonance width 
$\Gamma_n=\sum_\alpha\Gamma_{n,\alpha}$. The phase is given by 
$\EXP{\I\chi_\ab}=
\frac{\varphi_n(\alpha)^*\varphi_n(\beta)}
     {|\varphi_n(\alpha)\varphi_n(\beta)|}$.

\subsubsection*{Average charge}

Equation (\ref{qmoy}) gives~:
\be\label{avcwc}
\smean{\hat Q(t)} \simeq \sum_\alpha\int_0^\infty\D E\,f_\alpha(E)\,
\sum_n\frac{\Gamma_{n,\alpha}/\pi}{(E-E_n)^2+\Gamma_n^2}
=\sum_n \sum_\alpha\frac{\Gamma_{n,\alpha}}{\Gamma_n}
\left(\frac1\pi\arctan\frac{V_\alpha-E_n}{\Gamma_n}+\frac12\right)
\ee
where the sum over $n$ runs over the energies of the resonances (energies
of the isolated graph). $V_\alpha$ is the potential at contact $\alpha$.
This equation was derived in \cite{PeyDegDouMel02} by tracing out the lead's
degrees of freedom.
Since the average charge is the sum of contributions of the various levels,
we can consider only one level $E_n$.
If the level is below the potentials, $E_n<V_R<V_L$, the occupation of the
level is $1$.
On the other hand, if the level $E_n$ is between the potentials,
$V_R<E_n<V_L$, and far enough from them (on the scale $\Gamma_n$), it gives
a contribution ${\Gamma_{n,L}}/{\Gamma_n}$ to the average charge, which simply 
expresses that only the left scattering state is contributing to the occupation of 
the resonant level.

\subsubsection*{ Charge noise at finite frequency}
Let us now discuss the finite frequency structure of the charge noise
for weakly connected graphs. Equation (\ref{qfluc}) requires evaluating
\be
\left|
  \int_{\rm Graph}\D x\,
  \tilde\psi^{(\alpha)}_E(x)^*\tilde\psi^{(\beta)}_{E+\omega}(x)
\right|^2
\simeq
\left|
  \sum_n\frac1\pi
  \frac{\sqrt{E_n} w_\alpha w_\beta \varphi_n(\alpha)\varphi_n^*(\beta)}
       {(E-E_n-\I\Gamma_n)(E+\omega-E_n+\I\Gamma_n)}
\right|^2
\:.\ee
Let us keep only the diagonal elements in the double sum. 
This diagonal approximation  is valid in the limit of narrow resonances 
($\Gamma_n\ll|E_{n+1}-E_{n}|$) since the energies $E$ and $E+\omega$ are 
then compelled to be both in the neighbourhood of the level $E_n$.
Then the correlation appears as a sum of contributions of the different
energy levels~:
\be
S_{QQ}(\omega) \simeq\sum_n S^{(n)}_{QQ}(\omega)
\:.\ee
The contribution of the level $E_n$ reads~:
\be
S^{(n)}_{QQ}(\omega) = 2\pi
\sum_{\alpha,\beta} \int\D E\, f_\alpha(E)[1-f_\beta(E+\omega)]\:
\frac{\Gamma_{n,\alpha}/\pi}{(E-E_n)^2+\Gamma_n^2}\:
\frac{\Gamma_{n,\beta}/\pi}{(E+\omega-E_n)^2+\Gamma_n^2}
\:.\ee
Performing the integrals leads to
\be
S^{(n)}_{QQ}(\omega) = \frac{1}{2\pi}\frac{1}{1+{\omega^2}/{4\Gamma_n^2}}
\sum_{\alpha,\beta}\frac{\Gamma_{n,\alpha}\Gamma_{n,\beta}}{\Gamma_n^3}
\theta(\omega+V_\alpha-V_\beta)\:A(V_\alpha,V_\beta;\omega)
\ee
where $\theta(\omega)$ is the Heaviside function and
\bea
A(V_\alpha,V_\beta;\omega)&=&
\arctan\left(\frac{V_\alpha-E_n}{\Gamma_n}\right)
-\arctan\left(\frac{V_\beta-E_n}{\Gamma_n}\right)\nonumber\\
&&+\arctan\left(\frac{V_\alpha+\omega-E_n}{\Gamma_n}\right)
-\arctan\left(\frac{V_\beta-\omega-E_n}{\Gamma_n}\right)\nonumber\\
&&+\frac{\Gamma_n}{\omega}
\ln\frac{[(V_\alpha+\omega-E_n)^2+\Gamma_n^2]
         [(V_\beta-\omega-E_n)^2+\Gamma_n^2]}
        {[(V_\alpha-E_n)^2+\Gamma_n^2][(V_\beta-E_n)^2+\Gamma_n^2]}
\:.\eea

For the particular case of a two terminal geometry the noise reads
\bea
S^{(n)}_{QQ}(\omega) =
\frac{1}{2\pi\Gamma_n^3}\frac{1}{1+{\omega^2}/{4\Gamma_n^2}}
\left[
    \Gamma_{n,L}^2 \, \theta(\omega) \, A(V_L,V_L;\omega)
  + \Gamma_{n,R}^2 \, \theta(\omega) \, A(V_R,V_R;\omega)
\right.\nonumber\\
\left.
  + \Gamma_{n,L}\Gamma_{n,R} \, \theta(\omega+V) \, A(V_L,V_R;\omega)
  + \Gamma_{n,R}\Gamma_{n,L} \, \theta(\omega-V) \, A(V_R,V_L;\omega)
\right]
\:,\eea
where $V=V_L-V_R>0$ is the voltage drop.

\medskip

\noindent$\bullet$ {\bf Equilibrium case~:} $V_L=V_R=0$.

\begin{figure}[!ht]
\begin{center}
\includegraphics[scale=0.75]{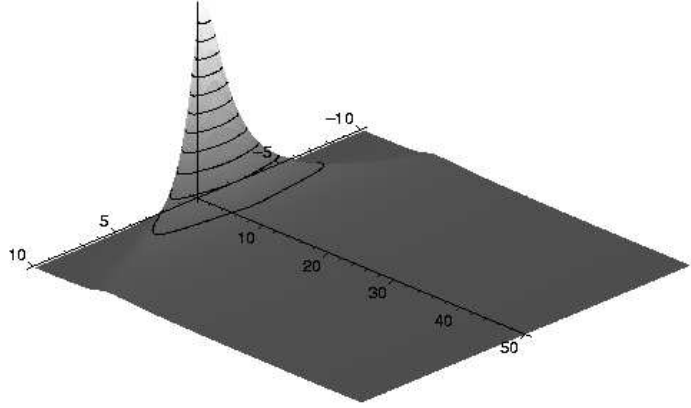}
\end{center}
\caption{Equilibrium noise in terms of $-10<\omega/\Gamma_n<10$ and
of the energy of the level $(-E_n)/\Gamma_n$ (this is equivalent to 
vary the energy or the chemical potential).
\label{fig:eqnoise}}
\end{figure}

\begin{center}
\includegraphics[scale=1]{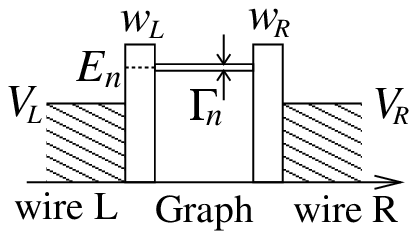}
\end{center}

\bea
S^{(n)}_{QQ}(\omega) =
\frac{1}{2\pi\Gamma_n}\frac{\theta(\omega)}{1+{\omega^2}/{4\Gamma_n^2}}
\left[
  \arctan\left(\frac{\omega-E_n}{\Gamma_n}\right)
  +\arctan\left(\frac{\omega+E_n}{\Gamma_n}\right)
\right.\nonumber\\
\left.
  +\frac{\Gamma_n}{\omega}
  \ln\frac{[(\omega-E_n)^2+\Gamma_n^2][(\omega+E_n)^2+\Gamma_n^2]}
          {[E_n^2+\Gamma_n^2]^2}
\right]
\:.\eea
We consider the case of narrow resonances where $\Gamma_n$ is the smallest
energy scale, then if the frequency is smaller than $|E_n|$, the 
contribution is zero, but if $\omega$ is sufficiently large to excite
an energy level $\omega\gtrsim|E_n|$, we get a contribution~:
\be\label{eqnoise}
S^{(n)}_{QQ}(\omega) \simeq
\frac{1}{2\Gamma_n}\frac{\theta(\omega)}{1+{\omega^2}/{4\Gamma_n^2}}
\times
\left\{
  \begin{array}{lll}
      0  & \mbox{ if } & \omega \lesssim |E_n| \\[0.25cm]
      1  & \mbox{ if } & \omega \gtrsim  |E_n|\:.
  \end{array}
\right.
\ee
Obviously, the transition between the two results is not sharp but occurs 
on a scale $\Gamma_n$.
Practically all the
noise power is concentrated at
low frequencies as shown on figure \ref{fig:eqnoise}.

Note that this contribution is independent of the fact that the level 
is occupied ($E_n<0$) or empty ($E_n>0$) since it is an even function
of $E_n$. 
At $V_L=V_R$, the low frequency charge noise can be understood
using a classical stochastic model describing the relaxation
process of an electron (or a hole) with lifetime $1/2\Gamma_n$.

\medskip

\noindent$\bullet$ {\bf Non equilibrium regime~:} $V_L\neq V_R$.

\noindent{\it Zero frequency limit.} 
Then only the term with $A(V_L,V_R;0)$ contributes~:
\bea
S^{(n)}_{QQ}(\omega=0)\simeq
\frac1\pi \frac{\Gamma_{n,R}\Gamma_{n,L}}{\Gamma_n^3}
\left\{ 
  \arctan\left(\frac{V_L-E_n}{\Gamma_n}\right) 
 +\arctan\left(\frac{E_n-V_R}{\Gamma_n}\right)
\right.\nonumber\\
\left.
  +\frac{\Gamma_n(V_L-E_n)}{(V_L-E_n)^2+\Gamma_n^2}
  +\frac{\Gamma_n(E_n-V_R)}{(E_n-V_R)^2+\Gamma_n^2}
\right\}
\:.\eea
Each energy level brings a contribution only if it is between the potentials
($V_R<E_n<V_L$) and far enough from them (on the scale $\Gamma_n$). In this 
case, the zero frequency charge noise is given by~:
\be
S^{(n)}_{QQ}(\omega=0) \simeq \frac{\Gamma_{n,R}\Gamma_{n,L}}{\Gamma_n^3}
\:.\ee
We recognize the factor $\Gamma_{n,R}\Gamma_{n,L}$ characteristic of
partition noise~: if one of the couplings vanishes ($\Gamma_{n,R}=0$ or
$\Gamma_{n,L}=0$) the occupation of the level is either $0$ or $1$ and
does not fluctuate. 
These fluctuations are a signature of the non equilibrium regime
of the mesoscopic circuit. 
The charge fluctuations at fixed time can be obtained by integrating the 
noise spectrum (\ref{cf1}).
Using the approximate expression (\ref{rwf}) in (\ref{cf2}) we obtain
\be
q_2 \simeq
\sum_n \sum_{\alpha,\beta}
\frac{\Gamma_{n,\alpha}\Gamma_{n,\beta}}{\Gamma_n^2}
\left(\frac1\pi\arctan\frac{V_\alpha-E_n}{\Gamma_n}+\frac12\right)
\left(\frac1\pi\arctan\frac{E_n-V_\beta}{\Gamma_n}+\frac12\right)
\ee
In the two leads case, only the levels between the two potentials bring the 
contribution~:
\be
q_2 \simeq \frac{\Gamma_{n,L}\Gamma_{n,R}}{\Gamma_n^2}
\:.\ee

\medskip

Note that this result cannot be simply infered from the current shot noise.
Near a resonance, the transmission probability through the graph
is $T(E)\simeq\frac{4\Gamma_{n,L}\Gamma_{n,R}}{(E-E_n)^2+\Gamma_n^2}$.
The average current in the lead is given by the Landauer formula
$\mean{I}=\frac{1}{2\pi}\int_{V_R}^{V_L}\D E\,T(E)$ whereas the current
and the shot noise by 
$S_{II}(\omega=0)=\frac{1}{2\pi}\int_{V_R}^{V_L}\D E\,T(E)(1-T(E))$
\cite{Les89,But90a,But92}.
If only one level $E_n$ lies between the two potentials, we obtain 
in this non linear regime \cite{BlaBut00}~:
$\mean{I}\simeq2\frac{\Gamma_{n,R}\Gamma_{n,L}}{\Gamma_n}$ and
$S_{II}(\omega=0)\simeq2\frac{\Gamma_{n,R}\Gamma_{n,L}}{\Gamma_n^3}
 (\Gamma_{n,R}^2+\Gamma_{n,L}^2)$.

\medskip

\noindent{\it Finite frequency noise.}
Let us choose the origin of the energies in such a way that~: $V_R=0$, $V_L=V>0$. 
Four energy scales must be considered~:
$E_n$, $\Gamma_n$, $V$ and $\omega$. Several regimes can be observed
according to the frequency. 
To help the discussion we neglect the smallest scale, supposed to 
be $\Gamma_n$, as we did above.

\noindent({\it i}) Let us first discuss the case of a fully occupied level~:
$E_n<V_R=0$. 

\begin{center}
\includegraphics[scale=1]{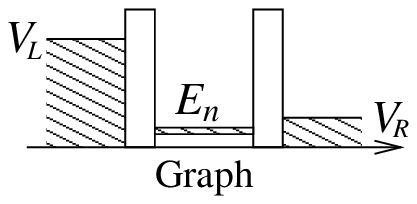}
\end{center}

\noindent
At small frequencies, correlations are roughly zero.
When $\omega$ reaches $V_R-E_n$, contributions from the second and third terms
appear, while all terms contribute for $\omega$ larger
than $V_L-E_n$. To summarize these three regimes~:
\be
S^{(n)}_{QQ}(\omega)\simeq
\frac{1}{2\Gamma_n}\frac{1}{1+{\omega^2}/{4\Gamma_n^2}}
\times
\left\{
  \begin{array}{lll}
    0                             & \mbox{ if } 
    & \omega \lesssim V_R-E_n \\[0.25cm]
    \frac{\Gamma_{n,R}}{\Gamma_n} & \mbox{ if } 
    & V_R-E_n \lesssim \omega \lesssim V_L-E_n
    \\[0.25cm]
    1                             & \mbox{ if } & V_L-E_n \lesssim \omega\:,
  \end{array}
\right.  
\ee
where we have factorized the equilibrium result.

\begin{figure}[!ht]
\begin{center}
\includegraphics[scale=0.75]{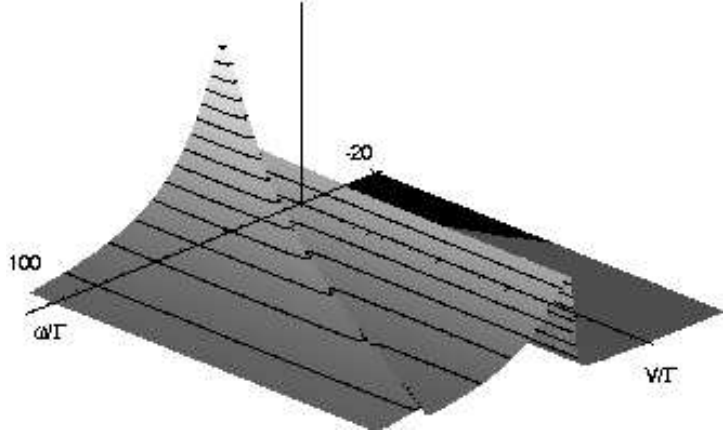}
\end{center}
\caption{Non equilibrium noise in terms of $-20<\omega/\Gamma_n<100$ and
         of the voltage drop $V/\Gamma_n$ for a fully populated level
         ($V_R=E_n + 50\Gamma_n$).\label{fig:neqnoise-deeplevel}}
\end{figure}

In the second regime $V_R-E_n \lesssim \omega \lesssim V_L-E_n$,
the energy $\omega$ is sufficient to excite the
state originating from the right reservoir but not from the left reservoir.
This is the origin of the ratio ${\Gamma_{n,R}}/{\Gamma_n}$. 
In the third regime, both reservoirs contribute to the noise. At fixed non zero 
bias voltage, this leads to a double peak structure
in terms of the $\omega$ which corresponds to the threshold for
creating electron-holes pairs involving the left and right leads (see figure 
\ref{fig:neqnoise-deeplevel}).
At small $V$ these two peaks tend to merge into a single more pronounced one.
At large $V$ the second peak occurs at a larger frequency and is less pronounced
because of the Lorentzian factor.

\noindent({\it ii}) Let us now discuss the case of an empty level~:
$E_n>V_L$. 

\begin{center}
\includegraphics[scale=1]{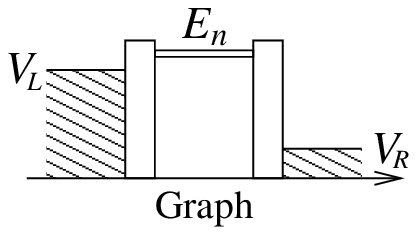}
\end{center}

\noindent
The physical picture can be obtained using a
hole picture. Three different regimes can also be distinguished~:
\be
S^{(n)}_{QQ}(\omega)\simeq
\frac{1}{2\Gamma_n}\frac{1}{1+{\omega^2}/{4\Gamma_n^2}}
\times
\left\{
  \begin{array}{lll}
      0                             & \mbox{ if } 
      & \omega \lesssim E_n-V_L \\[0.25cm]
      \frac{\Gamma_{n,L}}{\Gamma_n} & \mbox{ if } 
      & E_n-V_L \lesssim \omega \lesssim E_n-V_R
      \\[0.25cm]
      1                             & \mbox{ if } & E_n-V_R \lesssim \omega\:.
  \end{array}
\right.
\ee

\noindent({\it iii}) Finally we consider the case of a level between
the two potentials $V_R<E_n<V_L$.

\begin{center}
\includegraphics[scale=1]{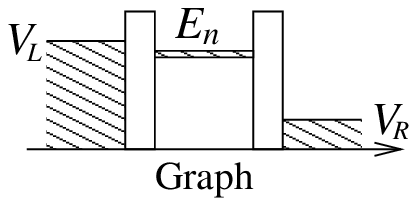}
\end{center}

\noindent
Considering the case where the level is closer to $V_L$ than to $V_R$,
we obtain
\be
S^{(n)}_{QQ}(\omega)\simeq
\frac{1}{2\Gamma_n}\frac{1}{1+{\omega^2}/{4\Gamma_n^2}}
\times
\left\{
  \begin{array}{lll}
    0              & \mbox{ if } & \omega \lesssim -E_n+V_R \\[0.25cm]
    \frac{\Gamma_{n,L}\Gamma_{n,R}}{\Gamma_n^2} 
                   & \mbox{ if } 
    & -E_n+V_R \lesssim \omega \lesssim -V_L+E_n \\[0.25cm]
    \frac{2\Gamma_{n,L}\Gamma_{n,R}}{\Gamma_n^2} 
                   & \mbox{ if } 
    & -V_L+E_n \lesssim \omega \lesssim V_L-E_n \\[0.25cm]
    \frac{2\Gamma_{n,L}\Gamma_{n,R}+\Gamma_{n,L}^2}{\Gamma_n^2}
                   & \mbox{ if } 
                   & V_L-E_n \lesssim \omega \lesssim E_n-V_R \\[0.25cm]
    1              & \mbox{ if } & E_n-V_R \lesssim \omega		       
  \end{array}
\right.
\ee

\begin{figure}[!ht]
\begin{center}
\includegraphics[scale=0.75]{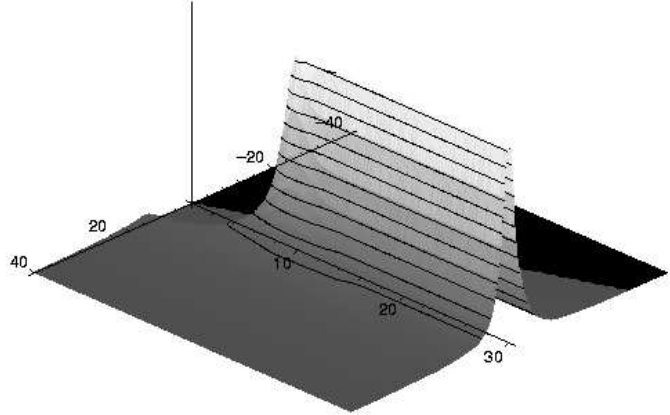}
\end{center}
\caption{Non equilibrium noise in terms of $-40<\omega/\Gamma_n<40$
         and of $V/\Gamma_n$. The right lead chemical potential
         is fixed to $-10\Gamma_n$. Case ({\it iii}) corresponds to 
         $V > 10\Gamma_n$ and exhibits an important low frequency
         noise whereas $V<10\Gamma_n$ corresponds to case ({\it ii}).
         \label{fig:noneqnoise1}}
\end{figure}

\begin{figure}[!ht]
\begin{center}
\includegraphics[scale=1]{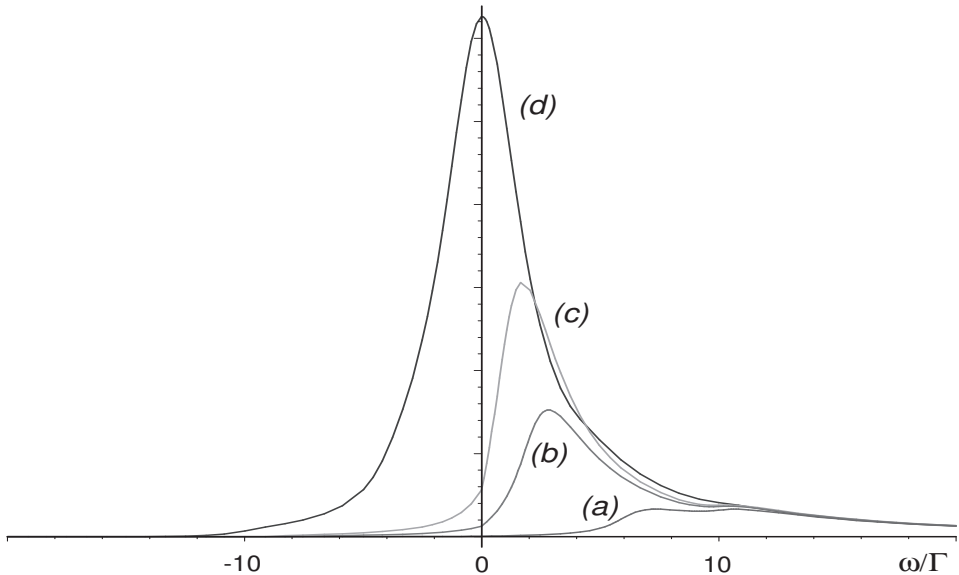}
\end{center}
\caption{Multiple plots of non equilibrium charge noise in terms of
         $-20<\omega/\Gamma_n<20$ for different values of $V/\Gamma_n$. The
         right lead chemical potential is fixed as on figure
         \ref{fig:noneqnoise1}. (a) $V/\Gamma_n=4$, (b) $V/\Gamma_n=8$, (c)
         $V/\Gamma_n=9$ and (d) $V/\Gamma_n=14$. \label{fig:noneqnoise2}}
\end{figure}

The fluctuation spectrum is symmetric in the interval centered
around $\omega=0$ with width of order $V$. The main
contribution to the noise appears at low frequency as can be seen
from figure \ref{fig:noneqnoise1}. The multiple plot
(figure \ref{fig:noneqnoise2}) shows how the low frequency peak develops
when $V_L$ crosses the energy level.

Interestingly, correlations are proportional to
the partition factor $\Gamma_{n,R}\Gamma_{n,L}$ only for small frequencies
$|\omega|\ll V$ (or large time scales $t\gg1/V$). 
For large frequencies $|\omega|\gg V$, the partition factor
does not appear. The high frequency part of the charge fluctuation spectrum is
insensitive to the fact
that the system is out of equilibrium. In this limit, the equilibrium
result (\ref{eqnoise}) is recovered.

\newpage 


\section{Currents inside the graph}

A possible way to probe a mesoscopic device is to attach some leads
to it, through which some currents are injected. Some information
can be extracted from transport or noise properties~:
average values and correlations of currents in the {\it external} leads.
All these properties can be related to the scattering matrix (see 
\cite{BlaBut00} for a review).
If one is now interested in local information on the system, like the
measurement of a persistent current, a natural way would be to introduce
some local probe. However, as we have recalled in the introduction,
local information can also be extracted from scattering properties.
Here, we investigate the currents in the {\it internal} wires, 
and show the relation to the scattering properties.
The starting point, exposed in the introduction, is the relation between the
current in a wire and the derivative of the scattering matrix with respect to
its conjugate variable, the flux in the wire. This idea comes from
\cite{AkkAueAvrSha91} and has been elaborated further in \cite{Tan01} to 
include derivation of correlations of the current density. 
Here,  we focus on the case of graphs in the context of which we will 
generalize these previous results to the non equilibrium situation.


\subsection{Current in a closed graph}

First we derive an expression for the current density in a closed graph.
It is convenient to introduce the spectral determinant of the Schr\"odinger 
operator
\be
S(\gamma) = \det(-\Delta+V(x)+\gamma) = \prod_n(E_n+\gamma)
\ee
where $\gamma$ is a spectral parameter. The set of $E_n$'s is the spectrum
of the Schr\"odinger operator on the graph. It was shown in
\cite{PasMon99,AkkComDesMonTex00,Des00,Des01} that the spectral determinant, 
which is the determinant of an unbounded operator, can be related to the
determinant of a finite size matrix. In the vertex approach, the formula (\ref{sd1})
involves a $V\times V$ matrix, whereas in the arc language the spectral 
determinant involves a 
$2B\times2B$ matrix~: $S(-E)\propto\det(1-QR)$ \cite{AkkComDesMonTex00,Des01}.
Note that for a closed graph, the vertex scattering matrix $Q$ has the same 
dimension as the bond scattering matrix.
We introduce the current density $j(E)$ associated to the states in the 
interval $[E,E+\D E[$.
The current density in the arc $a$ is~:
\be
j_a(E) = -\sum_n \delta(E-E_n)\drond{E_n}{\theta_a}
=\frac1\pi \im \drond{}{\theta_a}\ln S(-E+\I0^+)
\:,\ee
where $\theta_a$ is the magnetic flux along this arc.

\medskip

\noindent
{\it Example}~: Consider a closed ring of perimeter $l$ threaded by a flux 
$\theta$. Its spectral determinant is 
$S(\gamma)=\cosh(\sqrt{\gamma}\,l)-\cos(\theta)$ \cite{AkkComDesMonTex00}.
We write~: $\gamma=-k^2+\I0^+$, then 
$\cosh(\sqrt{\gamma}l)=\cos(kl)+\I0^+\sin(kl)$
and we get for the current density in the ring~:
\be
j(E)=-\sin\theta\,\sign(\sin kl)\,\delta(\cos kl-\cos\theta)
=\sum_n\delta(E-E_n)\,I_n
\:,\ee
with $I_n=-\partial_\theta E_n$ where $E_n=(2n\pi-\theta)^2/l^2$.


\subsection{Current in open graphs}

Now we consider a graph connected to infinite leads.

The current operator is 
\be
\hat J(x,t) = \frac1\I\left[
   \hat\psi^\dagger(x,t)\,\Dc_x\hat\psi(x,t) 
 - \Dc_x^*\hat\psi^\dagger(x,t)\,\hat\psi(x,t) 
\right]
\ee
where $\Dc_x=\D_x-\I A(x)$ is the covariant derivative.

We introduce the current matrix elements 
\be
j^{(\alpha,\beta)}_{\mu\nu}(E) = 
\bra{\tilde\psi^{(\alpha)}_E}\hat J(x,t)\ket{\tilde\psi^{(\beta)}_E}
\hspace{0.5cm}\mbox{for } x\in \mu\nu
\:\ee
which can be shown to be independent of the coordinate $x$ along the arc (only if the
two states have the same energy).
This matrix element can then be computed at the vertex $\mu$ ($x=0$)~:
\be
j^{(\alpha,\beta)}_{\mu\nu}(E)=
\frac1\I \left(
  \tilde\psi^{(\alpha)*}_{\mu}\,
  \Dc_x\tilde\psi^{(\beta)}_{(\mu\nu)}(\mu)
- \Dc_x^*\tilde\psi^{(\alpha)*}_{(\mu\nu)}(\mu)\,
  \tilde\psi^{(\beta)}_{\mu}
\right)
\:.\ee
The quantum and statistical average of the current operator in the 
arc $\mu\nu$ gives~:
\be\label{avc}
J_{\mu\nu}=\smean{\hat J(x\in \mu\nu,t)} = 
\sum_\alpha \int\D E\, f_\alpha(E)\,
j^{(\alpha,\alpha)}_{\mu\nu}(E)
\:.\ee
The correlations (unsymmetrized in time or frequency) between the currents 
in the arcs $\mu\nu$ and $\mu'\nu'$, defined as
\be\label{ccdef}
S_{J_{\mu\nu}J_{\mu'\nu'}}(\omega) = \int_{-\infty}^{+\infty}\D(t-t')\,
\left( \smean{\hat J(x,t)\,\hat J(x',t')}
       - \smean{\hat J(x,t)}\smean{\hat J(x',t')}
\right)\,\EXP{\I\omega(t-t')}
\:,\ee
for $x\in \mu\nu$ and $x'\in \mu'\nu'$.
They can be rewritten at zero frequency as~:
\be\label{cc1}
S_{J_{\mu\nu}J_{\mu'\nu'}}(\omega=0) 
= {2\pi}\sum_{\alpha,\beta}\int\D E\,
{f_\alpha(E)[1-f_\beta(E)]}\,
j^{(\alpha,\beta)}_{\mu\nu}(E)\,j^{(\beta,\alpha)}_{\mu'\nu'}(E)
\:.\ee

\subsubsection*{Vertex formulation}

Now we look for a relation between the current density and the scattering 
matrix. We start from the expression of the scattering state in the arc 
$\mu\nu$~:
\be\label{wf}
\psi_{(\mu\nu)}(x) = \EXP{\I \theta_{\mu\nu} x/l_{\mu\nu}}
\left( \psi_\mu\, f_{\mu\nu}(x)  
     + \psi_\nu\, \EXP{-\I\theta_{\mu\nu}} f_{\nu\mu}(x) \right)
\:.\ee
Then~:
\be
j^{(\alpha,\beta)}_{\mu\nu}=
\frac1\I \left(
- \tilde\psi_\mu^{(\alpha)*} \,
  \frac{\D f_{\mu\nu}}{\D x_{\mu\nu}}(\nu)\,\EXP{-\I\theta_{\mu\nu}} \,
  \tilde\psi_\nu^{(\beta)}
+ \tilde\psi_\nu^{(\alpha)*} \,
  \frac{\D f_{\nu\mu}}{\D x_{\nu\mu}}(\mu)\,\EXP{\I\theta_{\mu\nu}}\,
  \tilde\psi_\mu^{(\beta)}
\right)
\ee
where we have used the fact that the Wronskian of $f_{\mu\nu}(x)$ and 
$f_{\nu\mu}(x)$ reads~:
$\frac{\D f_{\nu\mu}}{\D x_{\mu\nu}}(\mu)
=-\frac{\D f_{\mu\nu}}{\D x_{\mu\nu}}(\nu)$. 
From the definition of the matrix $M$, we see that~:
\be
j^{(\alpha,\beta)}_{\mu\nu}=
-\frac{k}{\I} \left(
  \tilde\psi_\mu^{(\alpha)*} 
  \frac{\D M_{\mu\nu}}{\D \theta_{\mu\nu}}
  \tilde\psi_\nu^{(\beta)}
+ \tilde\psi_\nu^{(\alpha)*} 
  \frac{\D M_{\nu\mu}}{\D \theta_{\mu\nu}}
  \tilde\psi_\mu^{(\beta)}
\right)
\ee
Only the two elements $M_{\mu\nu}$ and $M_{\nu\mu}$ depend on the flux
$\theta_{\mu\nu}$, then~:
\be
j^{(\alpha,\beta)}_{\mu\nu}=
-\frac{k}{\I} \left( \tilde\Psi^\dagger \frac{\D M}{\D \theta_{\mu\nu}}
\tilde\Psi\right)_{\alpha\beta}
\:,\ee
where $\tilde\Psi$ is the $V\times L$-matrix that gathers the 
values of the $L$ stationary states at the $V$ vertices~:
$\tilde\Psi_{\mu\alpha}\equiv\tilde\psi^{(\alpha)}_\mu$. This
matrix is \cite{TexMon01,TexBut03}~:
\be
\tilde\Psi = \frac1{\sqrt{\pi k}}\frac{1}{M+W^{\rm T}W}W^{\rm T}
\:.\ee
We can rewrite the current density in terms of matrices $M$ and $W$~:
\be\label{cd1}
j^{(\alpha,\beta)}_{\mu\nu}
=-\frac{1}{\I\pi}
\left(
  W\frac{1}{-M+W^{\rm T}W}
  \frac{\D M}{\D \theta_{\mu\nu}}
  \frac{1}{M+W^{\rm T}W}W^{\rm T}
\right)_{\alpha\beta}
\:.\ee

Our aim is now to find the relation of this expression with the scattering 
matrix. We use the relation
$\frac{\D}{\D\eta}A(\eta)^{-1}
=-A(\eta)^{-1}\frac{\D A(\eta)}{\D\eta}A(\eta)^{-1}$
that gives the derivative of the inverse of
a square matrix $A(\eta)$ depending on a parameter $\eta$.
In the expression (\ref{RES2})
only $M$ depends on the fluxes, it follows that~:
\be
\frac{\D\Sigma}{\D \theta_{\mu\nu}}
=-2W\frac{1}{M+W^{\rm T}W}
\frac{\D M}{\D \theta_{\mu\nu}}
\frac{1}{M+W^{\rm T}W}W^{\rm T}
\:.\ee
If we multiply this expression by $\Sigma^\dagger$ from the left, it replaces 
the $M$ in the left fraction by $-M$. We conclude that the off-diagonal
elements of the current density reads~:
\be\label{cud}
j_{\mu\nu}^{(\alpha,\beta)}(E) = 
\frac1{2\I\pi}\left(
 \Sigma^\dagger\frac{\D\Sigma}{\D\theta_{\mu\nu}}\right)_{\alpha\beta}
\:.\ee
Note that this result is reminiscent to the one obtained by Taniguchi in
\cite{Tan01} who derived some relation between the scattering matrix
and the ``current density'', {\it i.e.} the diagonal elements ($\alpha=\beta$) 
of (\ref{cud}).

\subsubsection*{Arc formulation}

Let us now reformulate the previous demonstration in the arc language which allows 
to consider the most general case.

As explained in the introduction, the wave function on the bond $(\mu\nu)$
can be expressed as~:
\be
\psi_{(\mu\nu)}(x_{\mu\nu}) = 
B_{\mu\nu}\,\phi_{\mu\nu}(x_{\mu\nu})+B_{\nu\mu}\,\phi_{\nu\mu}(x_{\mu\nu})
\ee
where $\phi_{\mu\nu}(x_{\mu\nu})$ and $\phi_{\nu\mu}(x_{\mu\nu})$ are the 
left and right stationary scattering states for the bond potential 
$V_{(\mu\nu)}(x_{\mu\nu})$.
Using the expressions of these functions at the extremities of the bond
given in the introduction, we get the derivative of the wave function
at the vertex $\mu$~:
\be
\Dc_x\psi_{(\mu\nu)}(\mu)=
\I k\left[B_{\mu\nu}(1-r_{\mu\nu})-B_{\nu\mu}t_{\nu\mu}\right]
\:.\ee
We call $\tilde B^{(\alpha)}_{\mu\nu}$ the internal amplitude corresponding
to the stationary scattering state $\tilde\psi^{(\alpha)}_E(x)$. 
These amplitudes are obtained by solving the equations $B=QA$ and 
$A^{\rm int}=RB^{\rm int}$ with external amplitudes $A^{\rm ext}$ describing 
the injection of a plane 
wave on the lead arriving at vertex $\alpha$~: its components are
$\tilde A^{(\alpha){\rm ext}}_{i}
=\frac{1}{\sqrt{4\pi k}}\delta_{i,{\rm arc}\:\alpha}$
where ``${\rm arc}\:\alpha$'' designates the arc related to the lead issuing
from $\alpha$.
Then the amplitude
\be\label{amplarc}
\tilde B^{(\alpha)}_{\mu\nu}
=\frac{1}{\sqrt{4\pi k}}
\left[(1-Q^{\rm int}R)^{-1}\tilde Q^{\rm T}\right]_{\mu\nu,{\rm arc}\:\alpha}
\ee
is related to the matrix element between the internal arc $\mu\nu$ and the
external arc ``${\rm arc}\:\alpha$''.

After a little bit of algebra, we get for the current density on the arc
$\mu\nu$~:
\be
j^{(\alpha,\beta)}_{\mu\nu} = 2k\left[
  \tilde B^{(\alpha)*}_{\mu\nu} |t_{\mu\nu}|^2 \tilde B^{(\beta)}_{\mu\nu}
+ \tilde B^{(\alpha)*}_{\mu\nu} t_{\mu\nu}^*r_{\nu\mu}
     \tilde B^{(\beta)}_{\nu\mu}
- \tilde B^{(\alpha)*}_{\nu\mu} t_{\nu\mu}^*r_{\mu\nu}
     \tilde B^{(\beta)}_{\mu\nu}
- \tilde B^{(\alpha)*}_{\nu\mu} |t_{\nu\mu}|^2 \tilde B^{(\beta)}_{\nu\mu}
\right]
\:.\ee
We have used $t_{\mu\nu}^*r_{\nu\mu}=-r_{\mu\nu}^*t_{\nu\mu}$ coming from
the unitarity of $R$.
In the bond scattering matrix $R$, only the transmissions depend on the 
magnetic fluxes~: $t_{\mu\nu}\propto\EXP{\I\theta_{\mu\nu}}$. It follows
that, in the matrix $\frac{\D R^\dagger}{\D\theta_{\mu\nu}}R$, only the 
$2\times2$ block related to the arcs $\mu\nu$ and $\nu\mu$ is different 
from zero. It is given by~:
\be
\I\left(
  \begin{array}{cc}
    - |t_{\mu\nu}|^2 & -t_{\mu\nu}^*r_{\nu\mu} \\ 
    t_{\nu\mu}^*r_{\mu\nu} & |t_{\nu\mu}|^2
  \end{array}
\right)
\ee
Then it is straightforward to see that~:
\be
j^{(\alpha,\beta)}_{\mu\nu} = -2\I k\sum_{i,j}\tilde B^{(\alpha)*}_i
\left(\frac{\D R^\dagger}{\D\theta_{\mu\nu}}R\right)_{i,j}
\tilde B^{(\beta)}_j
\ee
where the sum over $i,j$ runs over the $2B$ internal arcs. Using the 
expression (\ref{amplarc}) for the amplitudes, we obtain~:
\bea
j^{(\alpha,\beta)}_{\mu\nu} &=& -\frac{1}{2\I\pi}
\left(
  \tilde Q^*(1-R^\dagger Q^{\rm int\:\dagger})^{-1} 
  \frac{\D R^\dagger}{\D\theta_{\mu\nu}}R\,
  (1-Q^{\rm int}R)^{-1}\tilde Q^{\rm T}
\right)_{\alpha,\beta} \\
&=& -\frac{1}{2\I\pi}
\left(
  \Sigma^\dagger
  \tilde Q (R^\dagger-Q^{\rm int})^{-1} \frac{\D R^\dagger}{\D\theta_{\mu\nu}}
  (R^\dagger-Q^{\rm int})^{-1}\tilde Q^{\rm T}
\right)_{\alpha,\beta} \\
&=& \frac{1}{2\I\pi}
\left(
  \Sigma^\dagger
  \tilde Q \frac{\D }{\D\theta_{\mu\nu}}
  (R^\dagger-Q^{\rm int})^{-1}
  \tilde Q^{\rm T}
\right)_{\alpha,\beta} 
= \frac{1}{2\I\pi}\left(
  \Sigma^\dagger\frac{\D\Sigma}{\D\theta_{\mu\nu}}
\right)_{\alpha,\beta} 
\:.\eea
We have recovered the formula (\ref{cud}) within the arc language
\footnote{
  We have used the relation
  $\Sigma^\dagger\tilde Q (R^\dagger-Q^{\rm int})^{-1}
  =\tilde Q^*(1-R^\dagger Q^{\rm int\:\dagger})^{-1}$,
  coming from the unitarity of the scattering matrices \cite{TexBut03}.
}.
This demonstrates that equation (\ref{cud}) applies to the most 
general situation, as expected.

\subsubsection*{Average current and current correlations in terms
                of the scattering matrix}

The average current can we written, as could have been guessed from the
general discussion of the introduction~:
\be\label{avc2}
J_{\mu\nu} = \sum_\alpha\int\D E\,f_\alpha(E)\,
\frac1{2\I\pi}\left(
 \Sigma^\dagger\frac{\D\Sigma}{\D\theta_{\mu\nu}}\right)_{\alpha\alpha}
\:.\ee

The correlations of currents at zero frequency rewrite in terms of scattering
matrix~:
\be\label{cc2}
S_{J_{\mu\nu}J_{\mu'\nu'}}(\omega=0) 
= - \frac{1}{2\pi}\sum_{\alpha,\beta}\int\D E\,
{f_\alpha[1-f_\beta]} 
\left(\Sigma^\dagger\frac{\D\Sigma}{\D\theta_{\mu\nu}}\right)_{\alpha\beta}
\left(\Sigma^\dagger\frac{\D\Sigma}{\D\theta_{\mu'\nu'}}\right)_{\beta\alpha}
\:.\ee
If $\mu\nu=\mu'\nu'$ this gives the noise of the persistent current.
At equilibrium, all potentials are equal $f_\alpha(E)=f(E)$ $\forall \alpha$, 
and we recover an expression reminiscent of the one given in \cite{Tan01}~:
\be\label{cc3}
S_{J_{\mu\nu}J_{\mu'\nu'}}(\omega=0) 
= \frac{1}{2\pi}\int\D E\,
f(E)[1-f(E)]\, 
\tr{\frac{\D\Sigma}{\D\theta_{\mu\nu}}
    \frac{\D\Sigma^\dagger}{\D\theta_{\mu'\nu'}}}
\:.\ee
In his work, Taniguchi identifies the contribution 
$\left(\frac{\D\Sigma}{\D\theta_{\mu\nu}}
    \frac{\D\Sigma^\dagger}{\D\theta_{\mu'\nu'}}\right)_{\alpha\alpha}$
of a given scattering state to this trace.
However we see that it is not sufficient to go back to
the expression (\ref{cc2}) describing the non equilibrium situation.


\subsection{Gauge invariance}

Since many formulae involve the fluxes along the wires, it is important to
discuss how a gauge transformation would affect them and to check that all
the measurable quantities are indeed gauge invariant. A gauge transformation
changes the vector potential according to $A(x)\to A(x)+\partial_x\chi(x)$, 
where $\chi(x)$ is a scalar function. 
The magnetic flux $\theta_{\mu\nu}$ along the arc $\mu\nu$ is then 
modified according to~:
\be
\theta_{\mu\nu} \hspace{0.5cm}\longrightarrow\hspace{0.5cm}
\theta'_{\mu\nu}=\theta_{\mu\nu}+\chi_\mu-\chi_\nu
\ee
where $\chi_\mu\equiv\chi(\mu)$ is the value taken by the function at the 
vertex $\mu$.
In the vertex approach, we immediatly see from its defintion that the matrix 
$M$ is changed as~:
\be
M_{\mu\nu} \hspace{0.5cm}\longrightarrow\hspace{0.5cm}
M'_{\mu\nu}=M_{\mu\nu}\:\EXP{\I\chi_\mu-\I\chi_\nu}
\:.\ee
We can write $M'={\cal U}M{\cal U}^\dagger$ where the diagonal unitary matrix  
reads~: ${\cal U}_\ab=\delta_\ab\EXP{\I\chi_\alpha}$. Since $W^{\rm T}W$ is
also diagonal it is clear that 
$(\pm M'+W^{\rm T}W)^{-1}={\cal U}(\pm M+W^{\rm T}W)^{-1}{\cal U}^\dagger$.
The scattering matrix changes in the same way~:
\be
\Sigma_\ab\hspace{0.5cm}\longrightarrow\hspace{0.5cm}
\Sigma'_\ab=\Sigma_\ab\:\EXP{\I\chi_\alpha-\I\chi_\beta}
\:.\ee
From (\ref{cd1}) or (\ref{cud}) we see that the matrix elements of the 
current operator pick up a phase through a gauge transformation
\be
j^{(\alpha,\beta)}_{\mu\nu}
\hspace{0.5cm}\longrightarrow\hspace{0.5cm}
j^{(\alpha,\beta)}_{\mu\nu}\:\EXP{\I\chi_\alpha-\I\chi_\beta}
\:.\ee
Nevertheless, the average current (\ref{avc}) and the correlations 
(\ref{cc1}) are gauge invariant, as they should.


\subsection{Weakly connected graphs}

As we did for the charge distribution, it is interesting to consider the case
of graphs weakly connected to the leads ($w_\alpha\to0$) for which interesting
results can be derived. The starting point is again the expression (\ref{rwf})
of the scattering state near a resonance (when $E$ is close to an eigenenergy 
of the isolated graph). Using this relation, the current density matrix 
element can be expressed as 
\bea\label{cdr}
j^{(\alpha,\beta)}_{\mu\nu}(E,E') 
&& \hspace{-0.4cm} 
 = \bra{\tilde\psi^{(\alpha)}_E}
  \hat J(x\in\mu\nu,t=0)\ket{\tilde\psi^{(\beta)}_{E'}}\\
&&  \hspace{-0.9cm}
  \APPROX{E,E'\sim E_n}\frac{\sqrt{E_n}}{\pi}\,
\frac{w_\alpha\varphi_n(\alpha)}{E-E_n-\I\Gamma_n}
\frac{w_\beta\varphi_n^*(\beta)}{E'-E_n+\I\Gamma_n} \:I_{\mu\nu}^n
\:,\eea
where $I_{\mu\nu}^n$ is the current in the arc $\mu\nu$ associated to the 
eigenstate $\varphi_n(x)$ of the isolated graph~:
\be
I_{\mu\nu}^n = -\I\varphi_n^*(x)\,\Dc_x\varphi_n(x) + {\rm c.c.}
\hspace{0.5cm}\mbox{for }x\in\mu\nu
\:.\ee
Note that in principle the current matrix element (\ref{cdr})
between scattering states of different energies depends on the coordinate 
$x$, however in the weak coupling limit, since the two scattering states 
are proportional to the same eigenstate $\varphi_n(x)$, the matrix element 
becomes $x$ independent. 

Equation (\ref{cdr}) shows that the calculation of the average current is very 
similar to the calculation of the charge (\ref{avcwc})~:
\be
J_{\mu\nu}
\simeq \sum_\alpha\int_0^\infty\D E\,f_\alpha\,
\sum_n\frac{I_{\mu\nu}^n\:\Gamma_{n,\alpha}/\pi}{(E-E_n)^2+\Gamma_n^2}
=\sum_n I_{\mu\nu}^n\sum_\alpha\frac{\Gamma_{n,\alpha}}{\Gamma_n}
\left(\frac1\pi\arctan\frac{V_\alpha-E_n}{\Gamma_n}+\frac12\right)
\:.\ee
The contribution of the resonant level can be written~:
\be\label{eq76}
J_{\mu\nu}^{(n)} \simeq I_{\mu\nu}^n \:\smean{\hat Q(t)}^{(n)}
\ee
where $\mean{\cdots}^{(n)}$ designates the contribution of the resonant 
level $n$.

Similarly we obtain for the correlations~:
\be
S^{(n)}_{J_{\mu\nu}J_{\mu'\nu'}}(\omega) \simeq 
I_{\mu\nu}^n\:I_{\mu'\nu'}^n\:S^{(n)}_{QQ}(\omega)
\:.\ee

For example, for a situation with two contacts with a potential drop $V$ and
only one resonant level contributing we get~:
\be
J_{\mu\nu} \simeq I_{\mu\nu}^n \frac{\Gamma_{n,L}}{\Gamma_n}
\ee
and
\be
S_{J_{\mu\nu}J_{\mu'\nu'}}(0) \simeq 
I_{\mu\nu}^n\:I_{\mu'\nu'}^n\: 
\frac{\Gamma_{n,L}\Gamma_{n,R}}{\Gamma_n^3}
\ee


\subsection{Example}

\begin{figure}[!ht]
\begin{center}
\includegraphics[scale=1]{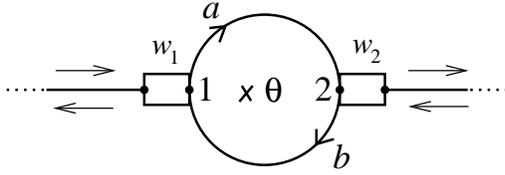}
\end{center}
\caption{A ring with two arms of lengths $l_a$ and $l_b$, threaded by a flux 
$\theta$ and coupled with two leads, with coupling parameters $w_{1,2}$. The
boxes represent the tunable couplings, with transmission amplitudes
$2w_{1,2}/(1+w_{1,2}^2)$ (see \cite{TexMon01}).\label{fig:ring}}
\end{figure}

Let us focus on the simple example of a ring  with two leads (see figure
\ref{fig:ring}).

The scattering matrix of the ring reads~:
\be
\Sigma = -1 + \frac{2}{\tilde S}
\left(\begin{array}{cc}
  \I w_1^2\sin kl + w_1^2w_2^2 s_a s_b  
& \I w_1 w_2 ( s_b\EXP{-\I\theta_a} + s_a\EXP{\I\theta_b} ) \\[0.2cm]
  \I w_2 w_1 ( s_b\EXP{\I\theta_a}  + s_a\EXP{-\I\theta_b} )
& \I w_2^2\sin kl + w_1^2w_2^2 s_a s_b
\end{array}\right)
\ee
where $\theta_a$ and $\theta_b$ are the fluxes of the two arcs and
$\theta=\theta_a+\theta_b$ the total flux threadening the ring.
We have denoted $s_{a,b}\equiv\sin kl_{a,b}$.
\be
\tilde S = s_as_b\det(M+W^{\rm T}W)
= 2(\cos\theta-\cos kl) + \I(w_1^2+w_2^2)\sin kl + w_1^2w_2^2 s_a s_b
\:\ee
is the modified spectral determinant.
The matrix involved in the current density in the arc $a$ is~:
\be
\Sigma^\dagger\frac{\D\Sigma}{\D\theta_a}
=\frac{2\sin\theta}{\tilde S}(1+\Sigma^\dagger)
+\frac{2w_1w_2 s_b}{\tilde S}
\left(\begin{array}{cc}
  -\Sigma_{21}^*\EXP{\I\theta_a} & \Sigma_{11}^*\EXP{-\I\theta_a}\\
  -\Sigma_{22}^*\EXP{\I\theta_a} & \Sigma_{12}^*\EXP{-\I\theta_a}\\
\end{array}\right)
\ee
from which we get the contribution of the scattering state 
$\tilde\psi^{(1)}(x)$ to the current density in the arc $a$~:
\be
j_a^{(1,1)}
=\frac1{2\I\pi}\left(\Sigma^\dagger\frac{\D\Sigma}{\D\theta_a}\right)_{11}
= \frac{2}{\pi|\tilde S|^2}
\left[
  -w_1^2\sin\theta\sin kl + w_1^2w_2^2(s_b^2+s_as_b\cos\theta)
\right]
\:.\ee

Let us now study the weak coupling limit $w_{1,2}\to0$.
Close to a resonance, we obtain~:
\be
j_a^{(1,1)}(k^2) \APPROX{k\sim k_n^\pm}
\mp \frac1l \frac{w_1^2}{w_1^2+w_2^2} \, 
\frac{\gamma/\pi}{(k-k_n^\pm)^2+\gamma^2}
\ee
where $\gamma=\frac{w_1^2+w_2^2}{2l}$. Integrating the contribution
of the resonance peak, we get~:
\be
\int_{k^\pm_n-\delta K}^{k^\pm_n+\delta K}\D k\,2k\,j_a^{(1,1)}(k^2)
\simeq \frac{w_1^2}{w_1^2+w_2^2}\,\frac{4\pi}{l^2}
\left(\mp n-\frac{\theta}{2\pi}\right)
\:.\ee
In the r.h.s we recognize the persitent current of the level of the isolated 
ring 
$-\drond{}{\theta}(k_n^\pm)^2
=-\drond{}{\theta}\left(\frac{2\pi n\pm\theta}{l}\right)^2$, 
multiplied by the ``relative weight''
$\frac{w_1^2}{w_1^2+w_2^2}$ of the scattering state $\tilde\psi^{(1)}(x)$,
as expected from (\ref{eq76}).


\subsection{Graphs with localized states}

In this section we discuss the consequence of the possible existence of 
localized states in certain graphs. These states are not probed by 
scattering, consequently their contributions to the current is not given by 
the expressions derived above.

\begin{figure}[!ht]
\begin{center}
\includegraphics[scale=1.25]{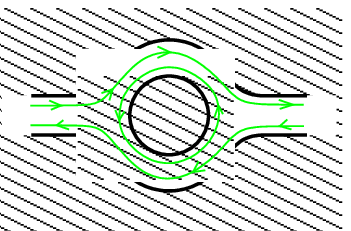}
\hspace{1cm}
\includegraphics[scale=1.25]{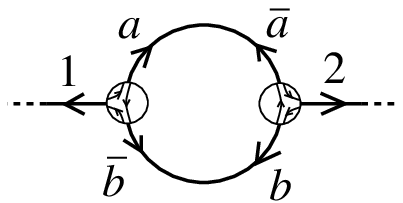}
\end{center}
\caption{{\it Left~:} A mesoscopic device in the regime of the IQHE with 
         an antidot at the middle. One edge state is open.
         {\it  Right~:} The graph that models this arrangement. 
         The scattering at the vertices is chiral.
	 \label{fig:QHring}}
\end{figure}

For the sake of simplicity our discussion will be focused on the 
example of a ring in the regime of the integer quantum Hall effect, 
with one edge state. The potential hill at the middle is called an antidot
(figure \ref{fig:QHring}, left).
This example must be thought more as a toy model to understand the idea of 
localized states in graphs, than as a reallistic model to describe 
current distribution in a quantum Hall device where the effect of screening
is important (the interested reader will find some discussion on the 
nature of edge currents and the role of screening in 
\cite{ChrBut96,KomHir96}).
The system can be modeled by a ring with chiral scattering at the vertices
(figure \ref{fig:QHring}, right).
The ring has two bonds, {\it i.e.} four internal arcs~: two arcs $a$ and $b$ 
carrying fluxes $\theta_a$ and $\theta_b$ and the two reversed arcs denoted
with a bar~: $\bar a$ and $\bar b$. The two leads are described by arcs 1 
and 2.

In the basis of arcs $\{a,\bar b,\bar a,b\,|1,2\}$, its vertex scattering 
matrix and bond scattering matrix are~:
\be
Q = \left(
\begin{array}{cccc|cc}
0 & 0 & 0 & 0 & 1 & 0 \\
1 & 0 & 0 & 0 & 0 & 0 \\
0 & 0 & 0 & 1 & 0 & 0 \\
0 & 0 & 0 & 0 & 0 & 1 \\\hline
0 & 1 & 0 & 0 & 0 & 0 \\
0 & 0 & 1 & 0 & 0 & 0 
\end{array}
\right)
=
\left(
\begin{array}{c|c}
Q^{\rm int} & \tilde Q^{\rm T} \\ \hline 
\tilde Q   & Q^{\rm ext}
\end{array}
\right)
\ee
The off diagonal blocks are not anymore transposed due to the breaking
of the time reversal symmetry at the vertices (chiral scattering), 
however we keep the same notation as above for simplicity since there is
no possible confusion.
On the other hand
\be
R = \left(
\begin{array}{cccc}
0&0&\EXP{\I kl_a-\I\theta_a}&0\\
0&0&0&\EXP{\I kl_b+\I\theta_b}\\
\EXP{\I kl_a+\I\theta_a}&0&0&0\\
0&\EXP{\I kl_b-\I\theta_b}&0&0
\end{array}
\right)
\:.\ee
We recall that $Q_{ij}$ is the transmission amplitude from arc $j$ to arc $i$
due to vertex scattering, and $R_{ij}$ describes bond scattering due to the
potential.

\vspace{0.25cm}

\noindent{\bf Scattering}.
The scattering matrix can be computed from $Q$ and $R$ with equation
(\ref{RES1}), however the result is obvious here, due to the absence of 
multiple scattering~:
\be
\Sigma=\left(
\begin{array}{cc}
0&\EXP{\I kl_b+\I\theta_b}\\
\EXP{\I kl_a+\I\theta_a}&0
\end{array}
\right)
\ee
The current density in the arc $a$ is given by the matrix~:
\be
\Sigma^\dagger\frac{\D\Sigma}{\D\theta_a}
=\left(\begin{array}{cc}\I&0\\0&0\end{array}\right)
\ee
From (\ref{cud}) we get
the contribution of the scattering state $\tilde\psi^{(1)}(x)$ to the 
current density in the arm $a$~: $j_a^{(1,1)}=\frac{1}{2\pi}$, whereas
the contribution of $\tilde\psi^{(2)}(x)$ obviously vanishes $j_a^{(2,2)}=0$,
since this latter scattering state does not send current into the arc $a$.

\vspace{0.25cm}

\noindent{\bf Localized states}.
We follow the discussion of \cite{Tex02,TexBut03}~: if localized states
are present, their discrete spectrum is given by solving 
$\det(R^\dagger-Q^{\rm int})=0$. Here, we see that the equation indeed 
possesses a set of solutions since
$\det(R^\dagger-Q^{\rm int})=\EXP{-\I kl}(\EXP{-\I kl}-\EXP{-\I\theta})$
where $\theta=\theta_a+\theta_b$. The spectrum of localized states 
is $k_n=(2\pi n+\theta)/l$ for $n\in\NN$ if $\theta\in[0,2\pi[$, since 
$k\geq0$ by convention. 
These states describe a clockwise motion of the electron in the loop 
of the graph (right part of figure \ref{fig:QHring}).
In the quantum Hall ring picture, they correspond to states whose
wave functions are localized on the edge of the antidot
(left part of figure \ref{fig:QHring}).
The current associated to the state of energy $k_n^2$ in the arc $a$ is 
$-\drond{}{\theta}k_n^2=-2(2n\pi+\theta)/l^2$.
Note that if one introduces some scattering on the bonds, the localized 
states are hybridized with the states of the continuum and 
the discrete part of the spectrum disappears.

The discrete spectrum also brings some contribution to the current in the 
arms of the ring, which cannot be obtained from the scattering properties.
Since the state $\tilde\psi^{(2)}$ does not contribute, the total average 
current in arm $a$ finally reads~:
\be
J_a = \int_0^\infty\D E\,f_1(E)\,j_a^{(1,1)}(E)
-\sum_{n=0}^\infty f_{\rm int}(k_n^2)\,\frac{2}{l^2}(2n\pi+\theta)
\ee
where $f_1(E)$ is the Fermi distribution for the lead $1$ and $f_{\rm int}(E)$
the Fermi distribution for the localized states inside the graph.

\vspace{0.25cm}

Now we can give the general expression for the current in the arc
$a$ for a graph with localized states. Since the discrete spectrum
of localized states $\{E_n\}$ is given by the equation
\be
0=\det(R^\dagger-Q^{\rm int})\propto\prod_n(E-E_n)
\ee
we have
\bea
J_{a} &=& \sum_\beta\int\D E\,f_\beta(E)\,
\frac1{2\I\pi}\left(
 \Sigma^\dagger\frac{\partial\Sigma}{\partial\theta_{a}}\right)_{\beta\beta}
\nonumber\\
&&+
\int\D E\,f_{\rm int}(E)
\frac1\pi 
\im \drond{}{\theta_a}
\ln 
  \left(\left.\det(R^\dagger-Q^{\rm int})
\right|_{E\to E+\I0^+}\right)
\:,\eea
where $f_{\rm int}(E)$ is the Fermi distribution associated to 
localized states.
The first term is the contribution of the scattering states whereas the 
second is the contribution of the localized states.


\section{Summary}

In this paper we have studied the two first cumulants of the charge of 
a graph connected to infinite wires, as well as the distribution of 
currents in the wires inside the graph. 
In particular, we have shown the relation with the scattering matrix,
allowing to study these quantities in an out of equilibrium situation,
when the graph is connected to wires put at different potentials.
We have obtained a formula for the average current and the current 
correlations that generalizes previous results 
known for the equilibrium situation \cite{AkkAueAvrSha91,Tan01}. 

We have also emphasized that the scattering matrix contains 
information only on the continuous part of the spectrum related to scattering 
states.
If some states remain localized in the graph, they give an additional 
contribution to the current not taken into account by the scattering 
approach.

We have considered the case of graphs weakly coupled to the leads. It is 
interesting to remark that the results obtained in this context are 
expected to be of much more generality than graphs, since the starting
point was to use an approximation of the scattering state near a resonant
level (\ref{rwf}), a form of great generality.
In particular, the contribution of the resonant level $n$ to the 
average of some quantity $X$ defined inside the graph reads
\be
\smean{\hat X(t)}^{(n)} \simeq X_n \:\smean{\hat Q(t)}^{(n)}
\:,\ee
where $X_n=\bra{\varphi_n}\hat X\ket{\varphi_n}$ is the expectation of 
$X$ in the eigenstate $\ket{\varphi_n}$ of the isolated system.
Similarly the contribution of the $n$-th resonant level to the correlations 
of two observables $X$ and $Y$ reads~:
\be
S^{(n)}_{XY}(\omega) \simeq X_n\:Y_n\:S^{(n)}_{QQ}(\omega)
\:.\ee
These results apply to a situation with narrow resonances 
($\Gamma_n\ll|E_{n+1}-E_n|$).
We repeat that we have not considered the effect of electron electron
interactions in this article (weakly connected devices with resonant tunneling
present in principle Coulomb blockade). It would be interesting to 
incorporate some effects of interaction. This could be already done in 
a mean field approximation to describe the effect of screening in the 
charge and current distribution following B\"uttiker's approach 
\cite{But93,ButPreTho93,ButThoPre94}.


\section*{Acknowledgements}

C.~T. would like to acknowledge fruitful discussions with
Markus B\"uttiker and also interesting remarks of Ken Ichiro Imura.


\begin{appendix}

\section{Relation between the different unsymmetrized correlators}
\label{app:unsymmetrized}

Let us consider $A$ and $B$ two hermitian operators associated to 
physical quantities in our system. There are two unsymmetrized 
correlation functions:
\bea
S_{A,B}(\omega) & = & \int_{-\infty}^{+\infty}\D\tau\,
\EXP{\I\omega \tau}\left(\langle A(t+\tau)\,B(t)\rangle -
\langle A(t+\tau)\rangle\,\langle B(t)\rangle\right)\\
\tilde{S}_{A,B}(\omega) & = & \int_{-\infty}^{+\infty}\D\tau\,
\EXP{\I\omega \tau}\left(\langle B(t)\,A(t+\tau)\rangle -
\langle B(t)\rangle\,\langle A(t+\tau)\rangle\right)
\:.\eea
These correlators do not depend on $t$ for a system in a stationary 
state\footnote{Glassy systems are an example for which this is not 
possible since the system never reaches a stationary state (weak 
ergodicity breaking).}. Note that, even if we consider out of equlibrium 
situations, only stationary states are considered in the present paper.

In this case, using time translation
invariance of one and two point correlation fonctions, we have~:
\be
\tilde{S}_{A,B}(\omega) = \int_{-\infty}^{+\infty}\D\tau\,
\EXP{\I\omega \tau}\left(\langle B(t-\tau)\,A(t)\rangle -
\langle B(t-\tau)\rangle\,\langle A(t)\rangle\right)
\ee
and this gives:
\be
\tilde{S}_{A,B}(\omega)=S_{B,A}(-\omega)
\:.\ee
When $A=B$, it shows that one unsymmetrized correlator 
determines the other one:
\be
\tilde{S}_{A,A}(\omega)=S_{A,A}(-\omega).
\:.\ee
Finally, note that the correlator ${S}_{A,A}(\omega)$ is real since the 
correlator in time obeys  $S_{A,A}(\tau)^*=S_{A,A}(-\tau)$.


\section{Structure of the stationary states near a resonance
         \label{app:resonance}}

If we consider a graph weakly coupled to the leads, we expect the 
stationary scattering states to be closely related to the eigenstates of 
the isolated graph. The purpose of the appendix is to demonstrate the 
precise relation.
The relations we will obtain are very reminiscent of the Hamiltonian 
approach of chaotic scattering \cite{MahWei69} (see also \cite{FyoSom96}). 
In this latter case some small couplings are introduced
between an isolated system and the leads, whereas we rather start from a
situation where the coupling can be arbitrary large and study the weak 
coupling limit to see how the properties of the isolated graph emerges
from its scattering properties.

Let us consider a graph ${\cal G}$, whose spectrum is supposed to be non 
degenerate for simplicity (the occurence of degeneracies leads to 
complications related to possible existence of localized states in the 
graph non probed by scattering \cite{Tex02,TexBut03}). In a first step we
describe how the eigenstates of the Schr\"odinger operator in the graph are 
constructed and in a second step we will establish the relation with
stationary states in the weak coupling limit.

\vspace{0.25cm}

\noindent{\bf Isolated graph}. 
The spectrum of the Schr\"odinger operator in the graph is given by the 
equation~: $S(\gamma) = 0$,
where $S(\gamma)=\prod_n(\gamma+E_n)$ is the spectral determinant, whose
construction is explained in \cite{PasMon99,AkkComDesMonTex00} for the 
case of free graphs, in \cite{Des00,Des00a} for graphs with potential
and in \cite{Des01} for graphs with general boundary conditions (more general 
than the continuity of the wave function at vertices). If the wave function
is continuous at vertices~:
\be\label{sd1}
S(\gamma)=\gamma^{V/2}\prod_{(\ab)}
\left(\frac{\D f_\ba}{\D x_\ab}(\alpha)\right)^{-1}\det M(\gamma)
\:.\ee
The product runs over all the bonds of the graph. We recall that
the functions $f_\ab(x)$ involved in $M(\gamma)$ are the solutions
of the Schr\"odinger equation on the bond
$[\gamma-\D_x^2+V_{(\ab)}(x)]f(x)=0$ for an energy $E=-\gamma$.
In general $S(\gamma) = 0$ possesses the same set of solutions as
\be
\det M(\gamma) = 0
\:.\ee
We do not discuss here the case where the sets of zeros of both equations do
not coincide, which is a little
bit pathological and would require to refine the following arguments. 
Let us however quote few examples of free graphs ($V(x)=0$) for which 
it is the case~:
the graph made of one line (in this case $\det M=1$ is independent of 
$\gamma$), the complete graph \cite{AkkComDesMonTex00,Tex02},...

The component of the wave function $\varphi_n(x)$ on the bond $(\ab)$ is~:
\be
\varphi_{n(\ab)}(x) = \EXP{\I A_\ab x}
\left( \varphi_{n,\alpha}\, f_\ab(x)  
     + \varphi_{n,\beta}\, \EXP{-\I\theta_\ab} f_\ba(x) \right)
\ee
where $\varphi_{n,\alpha}$ is the wave function at the vertex $\alpha$ and
$A_\ab=\theta_\ab/l_\ab$ the vector potential.
(Do not confuse the label $n$ of the eigenstate with the greek labels that
designate vertices).
If we gather the wave function at the nodes in the $V$-dimensional 
column vector $\varphi_n$, the eigenstate of energy $E_n$ is solution of 
\be
M(-E_n) \varphi_n = 0
\:.\ee

\noindent{\it Normalization}.
The normalization condition for the eigenstate reads~:
\be
\int_{\rm Graph}\D x\, \left|\varphi_n(x)\right|^2
=\sum_{(\ab)}\int_{0}^{l_\ab} \D x\, \left|\varphi_{n(\ab)}(x)\right|^2
=1
\:.\ee
If we use the following relations \cite{Des00}~:
\bea\label{relJ1}
\int_{0}^{l_\ab} \D x_\ab\, f_\ab(x_\ab)^2 &=& 
-\partial_\gamma \frac{\D f_\ab}{\D x_\ab}(\alpha) \\
\label{relJ2}
\int_{0}^{l_\ab} \D x_\ab\, f_\ab(x_\ab)f_\ba(x_\ab) &=& 
 \partial_\gamma \frac{\D f_\ab}{\D x_\ab}(\beta) 
\eea
we obtain
\bea
\int_{\rm Graph}\D x\, \left|\varphi_n(x)\right|^2
=\sum_{(\ab)}\left[
    \varphi_{n,\alpha}^* \,
    \partial_\gamma \left(-\frac{\D f_{\ab}}{\D x_{\ab}}(\alpha)\right) \,
    \varphi_{n,\alpha}\,
  + \varphi_{n,\alpha}^* \,
    \partial_\gamma \left(
      \frac{\D f_{\ab}}{\D x_{\ab}}(\beta)\,\EXP{-\I\theta_{\ab}}\right) \,
    \varphi_{n,\beta}
    \right.\nonumber\\ \left.
  + \varphi_{n,\beta}^*\,
    \partial_\gamma \left(
      \frac{\D f_{\ba}}{\D x_{\ba}}(\alpha)\,\EXP{\I\theta_{\ab}}\right) \,
    \varphi_{n,\alpha}\,
  + \varphi_{n,\beta}^*\,
    \partial_\gamma \left(-\frac{\D f_{\ba}}{\D x_{\ba}}(\beta)\right) \,
    \varphi_{n,\beta}
\right]
\eea
If we replace the sum over bonds by a sum over vertices, the matrix $M$
appears.
Finally, the normalization condition reads for the $V$-vector $\varphi_n$~:
\be\label{norm}
\varphi_n^\dagger \partial_\gamma[\sqrt\gamma M(\gamma)]
\varphi_n = 1
\:,\ee
where the spectral parameter is taken, after derivation, equal to the 
eigenenergy ${\gamma=-E_n-\I0^+}$.

\vspace{0.25cm}

\noindent{\bf Graph weakly connected to leads}.
When the graph is weakly coupled to leads ($w_\alpha\to0$) we expect that
the stationary state $\tilde\psi^{(\alpha)}_E(x)$ is proportional to the 
wave function of the isolated graph near the resonance $E\simeq E_n$~:
$\tilde\psi^{(\alpha)}_{E}(x)\propto\varphi_n(x)$ for $x\in{\cal G}$.
The question is how to recover precisely this relation from our formalism~?

\noindent{\it The resonance width}.
As a preliminary question, it is instructive to find an expression for the 
resonance widths. For this purpose, let us consider the determinant of the 
scattering matrix \cite{Tex02}~:
\be
\det\Sigma = (-1)^L \frac{\det(M - W^{\rm T}W)}{\det(M + W^{\rm T}W)}
\ee
and find an approximation near a resonance $E_n$.

For any fixed energy $E>0$, $M$ is an antihermitian matrix and can be 
written in terms of its purely imaginary eigenvalues $\I\lambda_\alpha(E)$ 
and its associate eigenvectors $v_\alpha(E)$~:
\be
M(-E) = \I \sum_{\alpha=1}^V \lambda_\alpha(E)\,v_\alpha(E)v_\alpha^\dagger(E) 
\:.\ee
The eigenvectors are normalized as $v_\alpha^\dagger v_\alpha=1$. 
If the energy $E$ is equal to the energy $E_n$ of an eigenstate of the 
isolated graph, one of the eigenvalues of $M$ is vanishing~: 
$\lambda_1(E_n)=0$. 
We suppose the spectrum of the isolated graph to be non degenerate. The 
eigenvector $v_1(E_n)$ coincides with the eigenstate ~: 
$v_1(E_n)=\nu_n^{-1}\varphi_n$~;  however these vectors are not normalized 
in the same way and differ in the multiplicative factor $\nu_n$.

Since $\det M(-E)$ is proportional to the spectral determinant and the 
spectrum supposed to be non degenerate, the eigenvalue $\lambda_1(E)$ 
behaves linearly  near the energy $E_n$~: $\lambda_1(E)\simeq(E-E_n)\beta_n$.
The normalization condition (\ref{norm}) reads~:
\be
-\varphi_n^\dagger \,
\partial_E\left(
  -\I\sqrt{E} \sum_{\alpha=1}^V \I\lambda_\alpha(E)\,
                v_\alpha(E)v_\alpha^\dagger(E)
\right)\bigg|_{E=E_n}\varphi_n = 1
\ee
then
\be
-\beta_n\sqrt{E_n}
 \varphi_n^\dagger v_1(E_n)\,v_1^\dagger(E_n)\varphi_n=1
\ee
We obtain the normalization constant~: $\nu_n=1/\sqrt{-k_n\beta_n}$
where $E_n=k_n^2$.

We now come back to $\det\Sigma$. In the weak coupling limit $w_\alpha\to0$
we can compute perturbatively the eigenvalues of $M\pm W^{\rm T}W$ to 
express the determinant~:
\bea
\det(M(-E)\pm W^{\rm T}W) &\APPROX{E\sim E_n}&
\prod_{\alpha=1}^V\left(
  \I\lambda_\alpha(E)\pm v_\alpha^\dagger(E) W^{\rm T}W v_\alpha(E)
\right) \\
&\simeq& \left(
  \I\beta_n\,(E-E_n)\pm v_1^\dagger(E_n) W^{\rm T}W v_1(E_n)
\right)\prod_{\alpha=2}^V \I\lambda_\alpha(E_n)
\eea
We can use the relation $v_1(E_n)=\sqrt{-k_n\beta_n}\,\varphi_n$ to 
get~:
\be
\det(M(-k^2)\pm W^{\rm T}W) \APPROX{k\sim k_n}
\left(k-k_n \pm \frac\I2 \varphi_n^\dagger W^{\rm T}W\varphi_n\right)
2\I\beta_nk_n\prod_{\alpha=2}^V \I\lambda_\alpha(k_n^2)
\:.\ee 
Then~:
\be
\det\Sigma \propto
\frac{k-k_n -\I\gamma_n}{k-k_n +\I\gamma_n}
\ee
where the resonance width in $k$-scale is~: 
$\gamma_n = \frac12\varphi_n^\dagger W^{\rm T}W\varphi_n=
\frac12\sum_{\alpha=1}^L w_\alpha^2|\varphi_{n,\alpha}|^2$.
This result is very satisfactory since it shows that
the lead $\alpha$ brings a contribution 
to the resonance width proportional to the transmission probability 
$w_\alpha^2$ between the graph and the lead\footnote{
  the transmission amplitude is $\frac{2w_\alpha}{1+w_\alpha^2}$ 
  for finite $w_\alpha$ \cite{TexMon01}.}
and to the probability density
$|\varphi_{n,\alpha}|^2\equiv|\varphi_n(\alpha)|^2$ associated to the 
eigenstate of the isolated graph, taken at the vertex $\alpha$ where the 
graph is connected.
In energy scale the resonance width reads~:
\be
\Gamma_n = 2k_n\gamma_n=\sqrt{E_n}\,\varphi_n^\dagger W^{\rm T}W\varphi_n
=\sum_{\alpha=1}^L \Gamma_{n,\alpha}
\ee
where
\be
\Gamma_{n,\alpha}=\sqrt{E_n}\, w_\alpha^2|\varphi_n(\alpha)|^2
\ee 
is the contribution of the lead $\alpha$.

\noindent{\it The wave function}.
We recall that the $V\times L$-matrix $\tilde\Psi$ that gathers the 
values of the $L$ scattering states at the $V$ vertices is \cite{TexMon01}~:
\be\label{wfv}
\tilde\Psi = \frac1{\sqrt{\pi k}}\frac{1}{M+W^{\rm T}W}W^{\rm T}
\:.\ee
We call $\psi^{(\alpha)}$ the $V$-column vector gathering the values of 
$\psi^{(\alpha)}(x)$ at the $V$ vertices~:
$\psi^{(\alpha)}=(\psi^{(\alpha)}_1,\cdots,\psi^{(\alpha)}_V)^{\rm T}$. 
The matrix $\Psi$ is obtained by gathering these $L$ column vectors~:
$\Psi = (\psi^{(1)},\cdots,\psi^{(L)})$.
In the weak coupling limit ($w_\alpha\to0$) and near the resonance $E_n$ 
we can keep only the contribution of the vanishing eigenvalue 
$\I\lambda_1(E)$ of $M$ to compute~:
\be
\frac{1}{M+W^{\rm T}W} \simeq 
\frac{1}{\I\beta_n(E-E_n) + v_1^\dagger(E_n) W^{\rm T}W v_1(E_n)}
v_1(E_n) v_1^\dagger(E_n)
\:.\ee
It follows that the scattering state at the vertex $\mu$ is~:
\be
\tilde\psi^{(\alpha)}_\mu=\tilde\Psi_{\mu\alpha}
\simeq\frac1{\sqrt{\pi k}}\frac{\I/2}{k-k_n +\I\gamma_n}
\varphi_{n,\mu}\,(\varphi_n^\dagger W^{\rm T})_\alpha
=\frac1{\sqrt{4\pi k}}
\frac{\I\,w_\alpha\,\varphi_{n,\alpha}^*}{k-k_n +\I\gamma_n}\varphi_{n,\mu}
\:.\ee
Since the vertex $\mu$ could be any point of the graph
because we have always the freedom to introduce
an additional vertex of weight $\lambda=0$ on any bond without changing the 
properties of the graph, we can rewrite more 
elegantly~:
\be
\tilde\psi^{(\alpha)}_{k^2}(x) \APPROX{k\sim k_n}
\frac{1}{\sqrt{4\pi k_n}}\,
\frac{\I w_\alpha \varphi_n^*(\alpha)}     
     {k-k_n+\I\gamma_n}\,
\varphi_n(x)
\ee
Or using energy scale~:
\be\label{pnar}
\tilde\psi^{(\alpha)}_E(x) \APPROX{E\sim E_n}
\frac{1}{\sqrt{\pi}}\,
\frac{\I E_n^{1/4} w_\alpha \varphi_n^*(\alpha)}{E-E_n+\I\Gamma_n}\,
\varphi_n(x)
\:.\ee

\noindent{\it The local density of states}. The contribution of the scattering
state $\tilde\psi^{(\alpha)}_E(x)$ to the off-diagonal LDoS 
$\bra{x}\delta(E-H)\ket{x'}$ is~:
\be
\tilde\psi^{(\alpha)}_E(x)\,\tilde\psi^{(\alpha)*}_E(x')
\APPROX{E\sim E_n}
\frac{\Gamma_{n,\alpha}/\pi}{(E-E_n)^2+\Gamma_n^2}\,
\varphi_n(x)\varphi_n^*(x')
\:.\ee
The LDoS is obtained by summing these contributions over $\alpha$.

\noindent{\it The scattering matrix}. The same discussion can be done to 
find an approximation for the scattering matrix. We obtain the well-known 
Breit-Wigner structure~:
\be
\Sigma_{\ab}(E) \APPROX{E\sim E_n} -\delta_{\ab} + 
\frac{2\I\sqrt{E_n}\, w_\alpha \varphi_n(\alpha)\,w_\beta \varphi_n^*(\beta)}
     {E-E_n+\I\Gamma_n}
\:.\ee


\section{Matrix elements of the charge operator\label{app:Smith}}

Our aim is to relate
\be\label{r1}
\rho^{(\alpha,\beta)}(E)=\int_{\rm Graph}\D x\,
\tilde\psi^{(\alpha)}_{E}(x)^*\,\tilde\psi^{(\beta)}_{E}(x)
=\sum_{(\mu\nu)}\int_0^{l_{\mu\nu}}\D x\,
\tilde\psi^{(\alpha)*}_{(\mu\nu)}(x)\,\tilde\psi^{(\beta)}_{(\mu\nu)}(x)
\ee
to the scattering matrix. The sum runs over the $B$ bonds.
The relation (\ref{Smith}) was proven for $\alpha=\beta$ in \cite{Tex02}
using a different method.

The computation of this integral follows exactly the lines of the one done
discussing the normalization of the states in the isolated graph. Then we
obtain~:
\be
\rho^{(\alpha,\beta)}(E)=\sum_{\mu,\nu}
\tilde\psi^{(\alpha)*}_\mu \, 
\partial_E \left(\I\sqrt{E}\, M_{\mu\nu}\right)\,
\tilde\psi^{(\beta)}_\nu
\:,\ee
that is~:
\bea
\rho^{(\alpha,\beta)}(E)&=& 
\left(
\tilde\Psi^\dagger \, 
\frac{\D}{\D E} \left(\I\sqrt{E}\, M\right)\, \tilde\Psi\right)_\ab\\
&=&-\frac{1}{2\I\pi}
\left(
W\frac{1}{-M+W^{\rm T}W}\left[
  2\frac{\D M}{\D E}+\frac1E M
\right]\frac{1}{M+W^{\rm T}W}W^{\rm T}
\right)_\ab
\:\eea
where we have used (\ref{wfv}).
From (\ref{RES2}) and 
\be
\frac{\D\Sigma}{\D E}=
-2W\frac{1}{M+W^{\rm T}W}\frac{\D M}{\D E}
\frac{1}{M+W^{\rm T}W}W^{\rm T}
\ee
we finally obtain the desired relation~:
\be\label{rab}
\rho^{(\alpha,\beta)}(E) = \frac{1}{2\I\pi}
\left(
  \Sigma^\dagger\frac{\D\Sigma}{\D E} + \frac{1}{4E}(\Sigma-\Sigma^\dagger)
\right)_\ab
\:.\ee

An alternative way to relate $\rho^{(\alpha,\beta)}(E)$ to derivative
of the scattering matrix is to introduce the variable conjugate to the
charge~:
a constant potential $U$ in the graph. The total potential now reads
$V(x)+U\,\theta_{{\cal G}}(x)$ where $\theta_{{\cal G}}(x)=1$ if
$x\in{\cal G}$ and $\theta_{{\cal G}}(x)=0$ if $x$ belongs to the leads.
In the presence of $U$, the function $f_\ab(x)$ involved in $M$ is solution 
of $[E+\D_x^2-V_{(\ab)}(x)-U]f_{\ab}(x)=0$. These functions are obtained by
a shift of the spectral parameter~: $f^U_\ab(x;E)=f^0_\ab(x;E-U)$.
It immediately follows that~:
\be
\rho^{(\alpha,\beta)}(E)= 
-\left(
\tilde\Psi^\dagger \, 
\frac{\D}{\D U} \left(\I\sqrt{E}\, M\right)\, \tilde\Psi\right)_\ab\\
\:.\ee
Using the same arguments as above we get~:
\be\label{r3}
\rho^{(\alpha,\beta)}(E)=
-\frac1{2\I\pi}\left( \Sigma^\dagger\frac{\D\Sigma}{\D U}\right)_\ab
\:.\ee
It is interesting to compare this relation with (\ref{rab})~: it 
shows that it is not similar to differentiate with respect to 
a constant potential $U$ or with respect to the energy since the potential
does not live in the wires. The difference however vanishes at high energy 
({WKB} limit).

A relation between the stationary states and 
the functional derivative of the scattering matrix \cite{But00}
\be\label{r2}
-\frac{1}{2\I\pi}
\left(
  \Sigma^\dagger\frac{\delta\,\Sigma}{\delta V(x)}
\right)_{\alpha\beta}
=\tilde\psi^{(\alpha)*}_E(x)\,\tilde\psi^{(\beta)}_E(x)
\ee
was proven for graphs \cite{TexBut03} where it is explained how it can be
computed with algebraic calculations only.
It follows that we can rewrite the equation (\ref{rab})~:
\be
-\int_{\rm Graph}\D x\,
  \frac{\delta\,\Sigma}{\delta V(x)}
=-\frac{\D\,\Sigma}{\D U} 
= \frac{\D\,\Sigma}{\D E} 
+ \frac{1}{4E}\left(\Sigma^2-1\right)
\:.\ee
The first equality, which is obtained by identification of 
(\ref{r1},\ref{r2}) with (\ref{r3}), is also a consequence of 
the definition of the functional derivative.

\end{appendix}



\end{document}